# Transient Impact Force and Spreading Dynamics of Droplets Impacting Spherical Surfaces

Mengqi Ye (叶孟琪)[a], Tianyou Wang (王天友)[a,b], Zhizhao Che (车志钊)[a,b,*]

[a] State Key Laboratory of Engines, Tianjin University, Tianjin, 300350, China

[b] National Industry-Education Platform of Energy Storage, Tianjin University, Tianjin, 300350, China

*Corresponding author: chezhizhao@tju.edu.cn

## Abstract

Droplet impacts on spherical surfaces are encountered in processes such as spray cooling, anti-icing, and spray deposition. The curvature of the impact target changes the spreading path of the droplet and the impact force, leading to more complex kinematic and transient force dynamics. This study employs a numerical model utilizing the volume of fluid approach to investigate the spreading, retraction, rebound and deposition characteristics of droplets impinging on spherical surfaces. The roles of wall wettability, spherical curvature, Weber number, and Ohnesorge number on droplet spreading and wall impact force are analyzed. The results show that the wall impact force exhibits two peaks in the complete rebound and partial rebound modes, whereas the second peak disappears in the complete deposition mode. Further parametric analysis shows that increasing the contact angle, i.e., decreasing wettability, reduces both the maximum spreading length and the contact time, and significantly changes the second peak of the impact force. Under complete rebound mode, the relationship between the contact time and the contact angle can be expressed as $t_c/t_\sigma = C_1\left[\left(1-\cos\theta_{eq}\right)^{-1/2} - 1/\sqrt{2}\right] + C_2$, where $C_1$ and $C_2$ are constants. Within the investigated curvature range, increasing the dimensionless curvature reduces the critical contact angle required for complete rebound. The dimensionless first peak of the impact force decreases with increasing the Weber number and finally approaches a constant, following a relationship of $F_{t1}^* = \beta_1 + \beta_2 We^{-1}$, where $\beta_1$ and $\beta_2$ are constants. Increasing the Ohnesorge number reduces the maximum spreading length, prolongs the contact time, and slightly increases the first peak of impact force.

## 1. Introduction

The impact of droplets onto solid surfaces is a multiphase flow phenomenon commonly encountered in nature and industrial processes. It involves complex processes such as rapid interfacial deformation, momentum transfer, energy dissipation, and solid-liquid interaction [1,2]. In applications including spray cooling, anti-icing and de-icing, spray deposition, particle coating, and additive manufacturing, the spreading, recoiling, deposition, rebound, and breakup of droplets after impact directly affect the liquid retention on solid surfaces, coverage uniformity, heat transfer efficiency, and the force response of the wall [3-6]. Therefore, an in-depth understanding of the kinematic evolution and dynamic response during droplet impact is essential for droplet deposition control, rapid detachment, and surface protection design.

Most fundamental studies have considered impact on planar surfaces. After impact, a droplet typically undergoes several successive stages, including initial contact, inertia-dominated spreading, maximum spreading, capillary retraction, and finally deposition, rebound, or splashing. These processes are governed by the competition among inertia, capillarity, and viscous effects [7-10]. In this context, the relative contributions of inertia, viscosity, and surface tension are commonly quantified using dimensionless groups, including the Weber, Reynolds, and Ohnesorge numbers. Early studies mainly focused on the morphological evolution of impacting droplets and established predictive correlations for the maximum spreading factor based on inertial-capillary balance or energy conservation. In recent years, researchers have gradually moved beyond single scaling laws with limited applicability toward multi-factor coupled predictions over broader parameter ranges. For example, some studies extended the applicability of maximum spreading models to highly viscous droplet impacts by modifying the viscous dissipation term in energy conservation models [11,12]. Other studies introduced machine learning methods based on large experimental datasets, in which droplet properties, impact conditions, and surface wettability were incorporated to estimate the maximum spreading factor [13,14]. Collectively, these investigations advance the understanding of the underlying mechanisms of droplet impact.

In practical and engineering applications, solid surfaces are often not ideal flat planes. Instead, they commonly have curved geometries. During droplet impingement, surface curvature alters the local contact geometry. As a consequence, the droplet no longer spreads radially along a flat surface, but moves along the arc direction or axial direction of the curved surface. Meanwhile, the local normal direction varies at different positions on the curved surface. The incident momentum is therefore redistributed between the normal and tangential directions. This redistribution affects droplet spreading, contact line motion, retraction path, rebound morphology, and the force response of the solid surface [15-18].

A sphere is a representative curved surface whose principal curvatures are identical at each location. Surface curvature primarily influences droplet impact through changes in the spreading length and film thickness. During meridional spreading over the spherical surface, the extent of contact-line motion is restricted by the sphere geometry. Therefore, the maximum spreading angle, spreading arc length, and lamella thickness are different from those on a flat surface. For conditions where the sphere diameter approximates the droplet diameter, the curvature effect becomes more pronounced. The droplet is more likely to extend along both sides of the sphere, and the local film thickness is also redistributed [19,20]. Increasing the Weber number or decreasing the surface contact angle usually enhances the spreading of the droplet on the spherical surface. A smaller size ratio between the sphere and the droplet can also lead to a larger spreading angle and a wider coverage range on the sphere [21,22]. For highly viscous droplets, viscous dissipation further weakens the inertial spreading of the liquid film. This weakens inertia-driven film expansion, thereby reducing the peak wetted area and modifying the temporal variations in film thickness and dynamic contact angle [23,24]. When the sphere is small relative to the droplet, the droplet is stretched along both sides of the sphere. The droplet may break up, or reconnect in the downstream region and then rupture. In this case, the droplet motion becomes more complex [25,26].

In addition to complete spherical surfaces, flat surfaces with hemispherical protrusions also provide an important reference for understanding how curvature regulates droplet impact. Relative to a smooth planar surface, the curved geometry modifies the initial configuration of the wetted region. The liquid can more easily bypass the protrusion and form an annular lamella. This further changes the subsequent spreading and retraction processes [27]. Upon impact on such hydrophobic surfaces, the conventional axisymmetric spreading and central retraction can be transformed into an annular spreading and annular retraction process. During retraction, the droplet forms an annular lamella and finally rebounds as an integral ring-like structure. Such behavior contributes to a notable reduction in the contact time [28,29].

Alongside the morphological evolution of the droplet, the transient loading exerted on the surface is also of considerable importance. The force generated upon impingement may result in surface erosion and damage [30-33]. For this reason, investigations into the spatiotemporal evolution of impact force in the course of impact are required. Compared with studies on droplet morphology, research on impact force remains relatively limited. Experimentally, the normal impact force is usually measured using piezoelectric sensors [34-36], whereas the measurement of shear stress requires high-speed stress microscopy [37]. With the development of measurement techniques, high-speed photoelastic tomography has been applied to impact force measurements [38]. This method enables non-contact and non-destructive two-dimensional full-field measurements of dynamic stress. Previous studies have shown that, in the initial phase of impact, the liquid at the bottom of the droplet is blocked by the solid surface. The fluid that initially moves in the normal direction rapidly turns into radial flow along the solid surface. Consequently, the impact force rises rapidly and attains a peak. Afterwards, as the droplet spreads along the wall, the normal momentum gradually decreases, leading to a more gradual attenuation of the impact force [39]. The early stage of droplet impact exhibits strong self-similar dynamics and follows a square-root time scaling law [35,40].

The total impact force is not the only quantity of interest. The distributions of local pressure and shear stress are also crucial to understanding droplet-impact dynamics. The total impact force can reflect the overall action of the droplet on the wall, but it cannot reveal the non-uniformity of local stresses. In contrast, the local stress distribution plays a decisive role in local surface erosion and material wear. Previous studies have shown that the maximum pressure at the early stage does not appear at the impact center. Instead, it occurs near the contact line and propagates outward as an annular high-pressure region [40,41]. The peak of shear stress is mainly associated with the growth of the near-wall boundary layer and is usually concentrated near the edge of the liquid film [37].

Beyond investigations of impact loading on planar surfaces, increasing attention has been devoted to the transient forces arising from droplet impingement on curved surfaces. Gu et al. [42] examined the impact behavior of droplets on single-hemispherical and double-hemispherical micro-structured surfaces. They found that the geometric size of the microstructures, wettability, and Weber number significantly changed the impact mode and the peak impact force. Xia et al. [43] employed a phase-field approach to characterize droplet impingement on a superhydrophobic convex hemisphere. At low Weber numbers, the air layer separating the droplet from the surface allowed contactless rebound. During this process, the force increased to a maximum and subsequently diminished, while the peak pressure remained approximately unchanged. At higher Weber numbers, the convex hemispherical structure reduced the maximum pressure during the initial impact stage. However, liquid convergence during retraction, central cavity collapse, and jet formation could induce a new pressure peak. For cylindrical surfaces, Aihara et al. [44] found that the transfer of normal momentum toward the wall became weaker as the surface curvature increased. This led to a reduction in the peak force exerted on the cylinder. Furthermore, an empirical correlation was developed to describe the connection between peak impact force and curvature radius. Ye et al. [45] performed numerical simulations of droplet impingement on a cylinder, focusing on the spreading behavior and transient loading. Their results revealed one force maximum in the deposition regime, whereas two separate maxima occurred during rebound. The first was governed mainly by the rapid transfer of the droplet's incident momentum to the surface, while the second was attributed to the reaction generated by inward liquid convergence and subsequent upward jetting during retraction. Chen et al. [46] quantified the transient loads produced by droplets striking a superhydrophobic cylinder using synchronized high-speed visualization and piezoelectric force sensing. Their results showed that increasing the surface curvature enhances the axial-azimuthal asymmetry, suppresses the formation of a Worthington jet, and consequently reduces both characteristic force peaks.

Although substantial progress has been made in understanding droplet impingement on solid surfaces, several unresolved issues remain. Existing studies mainly focus on the evolution of droplet morphology and the classification of different impact modes. In contrast, the force response of the solid surface, as well as the local pressure and shear stress distributions, has received limited attention. In contrast to planar surfaces, droplet impingement on curved surfaces remains insufficiently understood, especially the force response during droplet impact on spherical surfaces. Therefore, the present work employs numerical simulations to systematically examine the complete process of droplet impact on spherical surfaces from both kinematic and dynamic perspectives. For different impact modes, the droplet morphology, spreading factor, central film thickness, and energy conversion process are analyzed. The variations in the total impact force and its components during impact are then clarified. Finally, parametric analysis is implemented to account for various influencing factors including surface wettability, spherical curvature, Weber number, and Ohnesorge number. The influence of these parameters is evaluated in terms of the maximum spreading length, time to maximum spreading, contact time, peak impact force, and the occurrence time of the peak force.

## 2. Numerical method and validation

All numerical simulations are performed on the open-source platform OpenFOAM. The gas-liquid interface is resolved using the volume of fluid formulation available in the two-phase solver interFoam [47], whereas surface tension effects are modeled through the continuum surface force approach [48]. The liquid and gas phases are assumed to be incompressible and immiscible, and the flow is treated as laminar [49]. Dynamic contact angle is described by a modified Kistler model, which is implemented as a dynamic library for the solver.

### 2.1 Governing equations

The continuity, momentum, and phase-fraction transport equations used in the simulations are given as follows:

$$\nabla \cdot \mathbf{u}=0, \tag{1}$$

$$\frac{\partial \rho \mathbf{u}}{\partial t} + \nabla \cdot (\rho \mathbf{u}\mathbf{u}) = -\nabla p + \nabla \cdot \boldsymbol{\tau} + \rho \mathbf{g} + \mathbf{f}_\sigma , \tag{2}$$

$$\frac{\partial \alpha}{\partial t} + \nabla \cdot (\mathbf{u}\alpha) = 0 , \tag{3}$$

where $p$, $\mathbf{g}$, and $\mathbf{u}$ denote the pressure, gravitational acceleration, and velocity vector, respectively. The phase fraction is denoted by $\alpha$. The values $\alpha = 0$ and $\alpha = 1$ correspond to the gas and liquid phases, respectively, while $\alpha = 0.5$ represents the gas-liquid interface. The viscous stress tensor is expressed as $\boldsymbol{\tau} = 2\mu \mathbf{S} = \mu(\nabla \mathbf{u} + \nabla \mathbf{u}^{\mathrm{T}})$, where $\mathbf{S}$ is the strain-rate tensor. The surface tension source term is written as $\mathbf{f}_\sigma = \sigma\kappa\nabla\alpha$, which accounts for the effect of surface tension [48]. Here, $\sigma$ is the liquid surface tension, $\kappa = -\nabla \cdot \mathbf{n}$ is the interface curvature, and $\mathbf{n} = \nabla\alpha / |\nabla\alpha|$ is the unit normal vector of the interface. The density $\rho$ and dynamic viscosity $\mu$ of the fluid are evaluated from the phase fraction as:

$$\rho = \alpha\rho_{\mathrm{l}} + (1-\alpha)\rho_{\mathrm{g}} , \tag{4}$$

$$\mu = \alpha\mu_{\mathrm{l}} + (1-\alpha)\mu_{\mathrm{g}} , \tag{5}$$

where “l” and “g” are subscripts referring to the liquid and gas phases, respectively. To maintain a sufficiently sharp interface, an artificial convection term is introduced in the phase-fraction transport equation. This term compresses the phase fraction near the interface and reduces the interface smearing caused by numerical diffusion[50]:

$$\frac{\partial \alpha}{\partial t} + \nabla \cdot (\mathbf{u}\alpha) + \nabla \cdot \left[ \alpha(1-\alpha) c_\alpha |\mathbf{u}| \frac{\nabla\alpha}{|\nabla\alpha|} \right] = 0 , \tag{6}$$

where $\mathbf{u}_{\mathrm{c}} = c_\alpha |\mathbf{u}| (\nabla\alpha / |\nabla\alpha|)$ is the artificial interface-compression velocity, $c_\alpha$ is the compression coefficient, which is set to 1 in this study.

The governing equations are solved using the finite volume method. The temporal derivatives are discretized using the Euler scheme. The convective term in the momentum equation is discretized using the Gauss linear Upwind scheme, whereas the convection term in the phase-fraction equation is treated using the Gauss vanLeer scheme. The gradient and Laplacian terms are discretized using the Gauss linear and Gauss linear corrected schemes, respectively. Pressure-velocity coupling is handled using the PIMPLE algorithm, which provides improved numerical stability and efficiency [51]. To keep the phase fraction bounded within $0 \le \alpha \le 1$, the MULES [52] algorithm based on the flux correct transport (FCT) method [53] is used for the phase-fraction equation. The Gauss linear Upwind scheme is employed to reduce numerical diffusion in the velocity field, whereas the flux-limited Gauss vanLeer scheme is used to maintain interface sharpness and suppress numerical oscillations. Together with the MULES boundedness treatment and interface-compression term, these schemes improve the numerical resolution of thin liquid films and highly curved interfaces near the contact line. An adaptive time step strategy is adopted during the simulations, and the maximum Courant number is set to 0.2.

### 2.2 Dynamic contact angle model and its implementation

The contact angle $\theta_c$ strongly affects the spreading and retraction of an impacting droplet. In this study, the Kistler dynamic contact angle is adopted. Relative to a constant contact angle formulation, incorporating dynamic wetting provides a more accurate description of droplet motion, particularly in the presence of contact angle hysteresis. The Kistler dynamic contact angle model describes the variation in contact angle using the Hoffman function. The dynamic contact angle $\theta_{dy}$ is related to the fluid properties and the velocity of the three-phase contact line [54]

$$\theta_{\mathrm{dy}} = f_{\mathrm{Hoff}}\left[Ca_{\mathrm{cl}} + f_{\mathrm{Hoff}}^{-1}(\theta_{\mathrm{eq}})\right], \tag{7}$$

$$f_{\mathrm{Hoff}}(x) = \arccos\left\{1 - 2\tanh\left[5.16\left(\frac{x}{1+1.31x^{0.99}}\right)^{0.706}\right]\right\}, \tag{8}$$

where $\theta_{eq}$ denotes the static contact angle, and $f_{\mathrm{Hoff}}^{-1}(x)$ represents the inverse Hoffman function. The capillary number based on the contact line velocity is given by $Ca_{\mathrm{cl}} = \mu u_{\mathrm{cl}}/\sigma$, where $u_{cl}$ is the contact line velocity. It is evaluated locally from the wall-relative fluid velocity. Specifically, the wall-normal component of the relative velocity is first removed to obtain the velocity tangential to the wall. The interface normal is then projected onto the wall tangent plane to define the local direction normal to the contact line. The contact line velocity is obtained by projecting the wall-tangential velocity onto this direction. It is positive during spreading, negative during retraction, and zero when the contact line is stationary. For surfaces that are not perfectly smooth, contact angle hysteresis may exist. The effects of the advancing and receding contact angles, denoted by $\theta_{adv}$ and $\theta_{rec}$, respectively, should be considered. Equation (7) can then be written as [55]

$$\theta_{\mathrm{dy}} = \begin{cases} f_{\mathrm{Hoff}}\left[Ca_{\mathrm{cl}} + f_{\mathrm{Hoff}}^{-1}(\theta_{\mathrm{adv}})\right] & \text{for } u_{\mathrm{cl}} > 0, \\ \theta_{\mathrm{eq}} & \text{for } u_{\mathrm{cl}} = 0, \\ f_{\mathrm{Hoff}}\left[Ca_{\mathrm{cl}} + f_{\mathrm{Hoff}}^{-1}(\theta_{\mathrm{rec}})\right] & \text{for } u_{\mathrm{cl}} < 0. \end{cases} \tag{9}$$

Although the Kistler model can describe the variation in contact angle reasonably well, it may produce negative values when the receding angle is small and the retraction velocity is large [56], which lacks physical meaning. To overcome this limitation, considering that the contact angle during spreading and that during receding exhibit central symmetry [57,58], Eq. (9) can be modified as follows:

$$\theta_{\mathrm{dy}} = \begin{cases} f_{\mathrm{Hoff}}\left[Ca_{\mathrm{cl}} + f_{\mathrm{Hoff}}^{-1}(\theta_{\mathrm{adv}})\right] & \text{for } u_{\mathrm{cl}} > 0, \\ \theta_{\mathrm{eq}} & \text{for } u_{\mathrm{cl}} = 0, \\ \pi - f_{\mathrm{Hoff}}\left[\left|Ca_{\mathrm{cl}}\right| + f_{\mathrm{Hoff}}^{-1}(\pi - \theta_{\mathrm{rec}})\right] & \text{for } u_{\mathrm{cl}} < 0. \end{cases} \tag{10}$$

Contact angle hysteresis is neglected in the present study. Accordingly, the advancing contact angle, receding contact angle, and static equilibrium contact angle are assigned the same value in each simulation. Different advancing and receding angles could be specified to account for hysteresis, but this effect is not considered here.

The calculated contact angle is subsequently imposed as a boundary condition in the numerical model. In OpenFOAM, the prescribed contact angle is imposed through the boundary condition of the phase-fraction field. During the evaluation of the interface properties, the gas-liquid interface normal is first calculated from the phase-fraction gradient. For wall patches employing the contact angle boundary condition, the contact angle is then used to correct the interface normal. After the corrected interface normal is obtained, the wall-normal gradient of the phase fraction is updated. The phase-fraction boundary condition is subsequently reevaluated using this corrected gradient. Therefore, the prescribed dynamic contact angle is imposed through the coupled correction of the interface normal and the phase-fraction boundary gradient during the solution. In general, it is achieved by adjusting the normal vector of the gas-liquid interface near the wall [59]. Once the phase-fraction field has been updated and the interface reconstructed, an angle $\theta_{\mathrm{i}}$ is formed at the three-phase contact line. This angle is generally different from the prescribed contact angle. The reconstructed interface normal vector $\mathbf{n}_{\mathrm{i}}$ can be expressed as:

$$\mathbf{n}_{\mathrm{i}} = \mathbf{n}_{\mathrm{w}} \cos\theta_{\mathrm{i}} + \mathbf{n}_{\mathrm{t}} \sin\theta_{\mathrm{i}}. \tag{11}$$

The symbols $\mathbf{n}_{\mathrm{w}}$ and $\mathbf{n}_{\mathrm{t}}$ represent the wall-normal and wall-tangential unit vectors, respectively. The interface normal vector $\mathbf{n}_{\mathrm{dy}}$ corresponding to the prescribed contact angle $\theta_{\mathrm{dy}}$ can be written as:

$$\mathbf{n}_{\mathrm{dy}} = \mathbf{n}_{\mathrm{w}} \cos\theta_{\mathrm{dy}} + \mathbf{n}_{\mathrm{t}} \sin\theta_{\mathrm{dy}}. \tag{12}$$

Combining Eqs. (11) and (12), the corrected interface normal vector can be obtained as

$$\mathbf{n}_{\mathrm{dy}} = a\mathbf{n}_{\mathrm{w}} + b\mathbf{n}_{\mathrm{i}}, \tag{13}$$

$$a = \frac{\cos\theta_{\mathrm{dy}} - \cos\theta_{\mathrm{i}} \cos\left(\theta_{\mathrm{i}} - \theta_{\mathrm{dy}}\right)}{1 - \cos^2\theta_{\mathrm{i}}}, \tag{14}$$

$$b = \frac{\cos\left(\theta_{\mathrm{i}} - \theta_{\mathrm{dy}}\right) - \cos\theta_{\mathrm{i}} \cos\theta_{\mathrm{dy}}}{1 - \cos^2\theta_{\mathrm{i}}}. \tag{15}$$

Thus, the corrected gas-liquid interface can be determined from $\mathbf{n}_{\mathrm{w}}$, $\mathbf{n}_{\mathrm{i}}$, $\theta_{\mathrm{i}}$, and $\theta_{\mathrm{dy}}$. In this way, the prescribed contact angle condition is imposed at the wall.

### 2.3 Energy calculation

During the impact process, the energy conversion mainly involves four parts: kinetic energy, gravitational potential energy, surface energy, and viscous dissipation. The energy components are evaluated using formulations that have been widely adopted in previous droplet-impact studies[60]. The kinetic energy $E_{\mathrm{k}}$ is calculated as:

$$E_{\mathrm{k}} = \frac{1}{2}\rho_{\mathrm{l}}\int_V \mathbf{u}_{\mathrm{l}}^2 dV, \tag{16}$$

where, $V$ denotes the liquid phase region, defined as the region where the phase fraction exceeds 0.5. The gravitational potential energy $E_{\mathrm{g}}$ is given by:

$$E_{\mathrm{g}} = \rho_{\mathrm{l}} g \int_V z dV. \tag{17}$$

The surface energy $E_\sigma$ consists of the interfacial energies of the solid-gas, solid-liquid, and gas-liquid interfaces, denoted by $E_{\mathrm{sg}}$, $E_{\mathrm{sl}}$ and $E_{\mathrm{lg}}$, respectively. It can be expressed as:

$$E_\sigma = E_{\mathrm{sg}} + E_{\mathrm{sl}} + E_{\mathrm{lg}} = \sigma\left(S_{\mathrm{gl}} - S_{\mathrm{sl}}\cos\theta_{\mathrm{eq}}\right), \tag{18}$$

where $S_{\mathrm{gl}}$ and $S_{\mathrm{sl}}$ are the gas-liquid and solid-liquid interfacial areas, respectively. The viscous dissipation during spreading and retraction $E_{\mathrm{diss}}$ is expressed as [61]

$$E_{\mathrm{diss}} = \int_0^t \int_V \phi dV dt, \tag{19}$$

where the viscous dissipation function $\phi$ is defined as[62,63]

$$\begin{aligned}\phi = 2\mu_{\mathrm{l}}(\mathbf{S}:\mathbf{S}) = 2\mu_{\mathrm{l}}\Bigg\{&\left(\frac{\partial u_{\mathrm{x}}}{\partial x}\right)^2 + \left(\frac{\partial u_{\mathrm{y}}}{\partial y}\right)^2 + \left(\frac{\partial u_{\mathrm{z}}}{\partial z}\right)^2 \\ &+ \frac{1}{2}\left[\left(\frac{\partial u_{\mathrm{y}}}{\partial x} + \frac{\partial u_{\mathrm{x}}}{\partial y}\right)^2 + \left(\frac{\partial u_{\mathrm{z}}}{\partial y} + \frac{\partial u_{\mathrm{y}}}{\partial z}\right)^2 + \left(\frac{\partial u_{\mathrm{x}}}{\partial z} + \frac{\partial u_{\mathrm{z}}}{\partial x}\right)^2\right]\Bigg\}.\end{aligned} \tag{20}$$

The instantaneous total energy of the impacting droplet, denoted by $E_{\mathrm{tot}}$, is then written as

$$E_{\mathrm{tot}} = E_{\mathrm{k}} + E_{\mathrm{g}} + E_\sigma. \tag{21}$$

According to energy conservation, the initial total energy of the droplet $E_{\mathrm{tot\text{-}0}}$ is defined as

$$E_{\mathrm{tot\text{-}0}} = E_{\mathrm{k\text{-}0}} + E_{\mathrm{g\text{-}0}} + E_{\sigma-0} = E_{\mathrm{tot}} + E_{\mathrm{diss}}, \tag{22}$$

where $E_{\mathrm{k\text{-}0}}$, $E_{\mathrm{g\text{-}0}}$, and $E_{\sigma\text{-}0}$ are the droplet's kinetic, gravitational potential, and surface energies evaluated at the initial moment, respectively.

### 2.4 Calculation of the impact force on the wall

Throughout the droplet impact process, the droplet exerts a force on the wall. In the present study, the reported wall impact force specifically refers to the liquid-induced force acting on the wall, while the direct mechanical load exerted by the surrounding gas on the exposed solid surface is neglected. Since the droplet impact considered in this study is axisymmetric, the resultant force acting on the wall is zero in the horizontal direction and only varies in the vertical direction. Taking the droplet as the object of analysis and neglecting the force exerted by the surrounding gas during impact, the droplet is subjected to gravitational force and the force by the wall in the vertical direction. The latter is denoted by $F'_{\mathrm{tot}}$. Here, the vertically downward direction is defined as positive. The same sign convention is adopted for all force components discussed below. The droplet impulse, denoted by *I*, is expressed as

$$I(t) = \Delta P = \int_0^t \left[-F'_{\mathrm{tot}}(t) + G\right] dt . \tag{23}$$

Following previous studies[64], the direct force exerted by the surrounding gas is neglected in the droplet momentum balance, as its contribution is much smaller than the liquid inertial force under the present conditions. Therefore, the total impact force acting on the wall, $F_{\mathrm{tot}}$, is equal in magnitude but opposite in direction to $F'_{\mathrm{tot}}$. Its magnitude is given as

$$F_{\mathrm{tot}}(t) = F'_{\mathrm{tot}}(t) = -\frac{dI(t)}{dt} + G , \tag{24}$$

where $G = \rho_{\mathrm{l}} g \pi D_{\mathrm{d}}^3 / 6$ is the weight of the droplet. The change in droplet momentum $\Delta P$ is expressed as:

$$\Delta P = P(t) - P(0) = \int_V \rho_{\mathrm{l}} u' dV - \int_V \rho_{\mathrm{l}} u_0 dV , \tag{25}$$

where $u'$ is the droplet velocity at time *t*. In Eq. (25), the time derivative of the impulse $dI/dt$ is evaluated using the central difference scheme in the numerical simulation. The impact force evaluated using Eq. (24) is derived directly from the integral momentum balance of the droplet. The proposed method has been employed in previous studies to determine or interpret the transient force generated during droplet impact [64]. The validation of the transient impact force is provided in the Supplementary Material (Fig. S3).

To further analyze the force acting on the wall, $F_{\mathrm{tot}}$ is decomposed into the pressure force $F_{\mathrm{p}}$, viscous force $F_{\mathrm{v}}$, and capillary force at the contact line $F_{\mathrm{cl}}$. The pressure and viscous forces are obtained by integrating the pressure and viscous tractions over the solid-liquid interface, respectively. The capillary force is obtained by integrating the surface tension along the contact line:

$$F_{\mathrm{p}} = \int_{S_{\mathrm{sl}}} (p - p_0)(\mathbf{n}_{\mathbf{wn}} \cdot \mathbf{z}) dS_{\mathrm{sl}} , \tag{26}$$

$$F_{\mathrm{cl}} = -\int_{L_{\mathrm{cl}}} \sigma \left[ \sin\theta_{\mathrm{c}} (\mathbf{n}_{\mathbf{wn}} \cdot \mathbf{z}) + \cos\theta_{\mathrm{c}} (\mathbf{n}_{\mathbf{wt}} \cdot \mathbf{z}) \right] dL_{\mathrm{cl}} , \tag{27}$$

$$F_{\mathrm{v}} = \int_{S_{\mathrm{sl}}} (\boldsymbol{\tau} \cdot \mathbf{n}_{\mathbf{wn}}) \cdot \mathbf{z} dS_{\mathrm{sl}} , \tag{28}$$

where $p$ is the local pressure acting on the solid-liquid interface and $p_0$ represents the atmospheric pressure. $\mathbf{n}_{\mathbf{wn}}$ and $\mathbf{n}_{\mathbf{wt}}$ are the unit normal vectors of the wall pointing toward the fluid side and the unit tangential vector of the wall pointing toward the top of the sphere, respectively. To further verify the consistency of the impact force calculation, the total wall impact force obtained from the droplet momentum balance is compared with the sum of the pressure, viscous, and capillary force contributions. As shown in Fig. S4 of the Supplementary Material, the two force histories exhibit close agreement throughout the impact process, including both characteristic force peaks, confirming the numerical consistency of the force calculation methods.

### 2.5 Physical model and validation

The physical model includes a droplet with a diameter of $D_{\mathrm{d}}$ impacting a spherical surface with a diameter of $D_{\mathrm{p}}$ at an impact velocity of $U_{\mathrm{i}}$, as shown in Fig. 1(a). After the impact, the droplet spreads symmetrically along the spherical surface. The spreading arc length is defined as $D_{\mathrm{s}}$, as indicated by the red arc in Fig. 1(b). The spreading angle is defined as $\omega = 2D_{\mathrm{s}}/D_{\mathrm{p}} = 4R_{\mathrm{s}}/D_{\mathrm{p}}$ , and the liquid film thickness at the center is denoted by $H_{\mathrm{fc}}$. The gravitational acceleration is set to 9.8 m/s$^2$.

In the present study, the droplet impacts the spherical surface along its central axis. Since the investigated cases are restricted to relatively low Weber numbers and dimensionless curvatures, no liquid-sheet breakup, splashing, or pronounced three-dimensional instability occurs within the parameter range considered. The droplet impacts normally along the central axis of the spherical surface, the spreading and retraction processes are predominantly axisymmetric. The liquid column breakup observed in the partial rebound mode mainly results from axial necking and pinch-off, whereas non-axisymmetric instabilities cannot be captured by the present model. Similar axisymmetric wedge formulations have been employed in previous numerical studies of droplet impact [55]. The dominant impact dynamics therefore remain approximately axisymmetric, and a rotationally symmetric wedge-shaped computational domain is employed in OpenFOAM. The wedge angle is 2°. The sector-shaped computational domain spans 90° in the meridional plane and extends radially to $5D_{\mathrm{d}}$, as shown in Fig. 1(c). The droplet is initially placed above the spherical surface with a distance of 0.5 mm between the droplet bottom and the sphere apex. A hexahedral structured mesh is adopted. The independence of the numerical results from the computational domain size and wedge angle has been further verified, as detailed in the Supplementary Material (Fig. S1). The spherical surface is treated as a no-slip wall, while the other boundaries are assigned pressure outlet boundary conditions. Mesh independence is assessed by comparing the temporal evolution of the spreading length $D_s$ at different mesh resolutions, as shown in Fig. 1(d). The mesh spacing in the radial direction is uniformly 10 μm, while the circumferential cell size is approximately 10 μm at the inner curved boundary and gradually increases with radial distance. Therefore, 10 μm refers to the minimum local mesh size rather than a uniform cell size throughout the domain. When the mesh size is reduced to 10 μm, $D_s$ no longer changes noticeably with further mesh refinement. Both the magnitudes and occurrence times of the first and second force peaks show only minor differences between the two finest meshes (Supplementary Material Fig. S2). Therefore, this mesh size is used in the subsequent simulations.

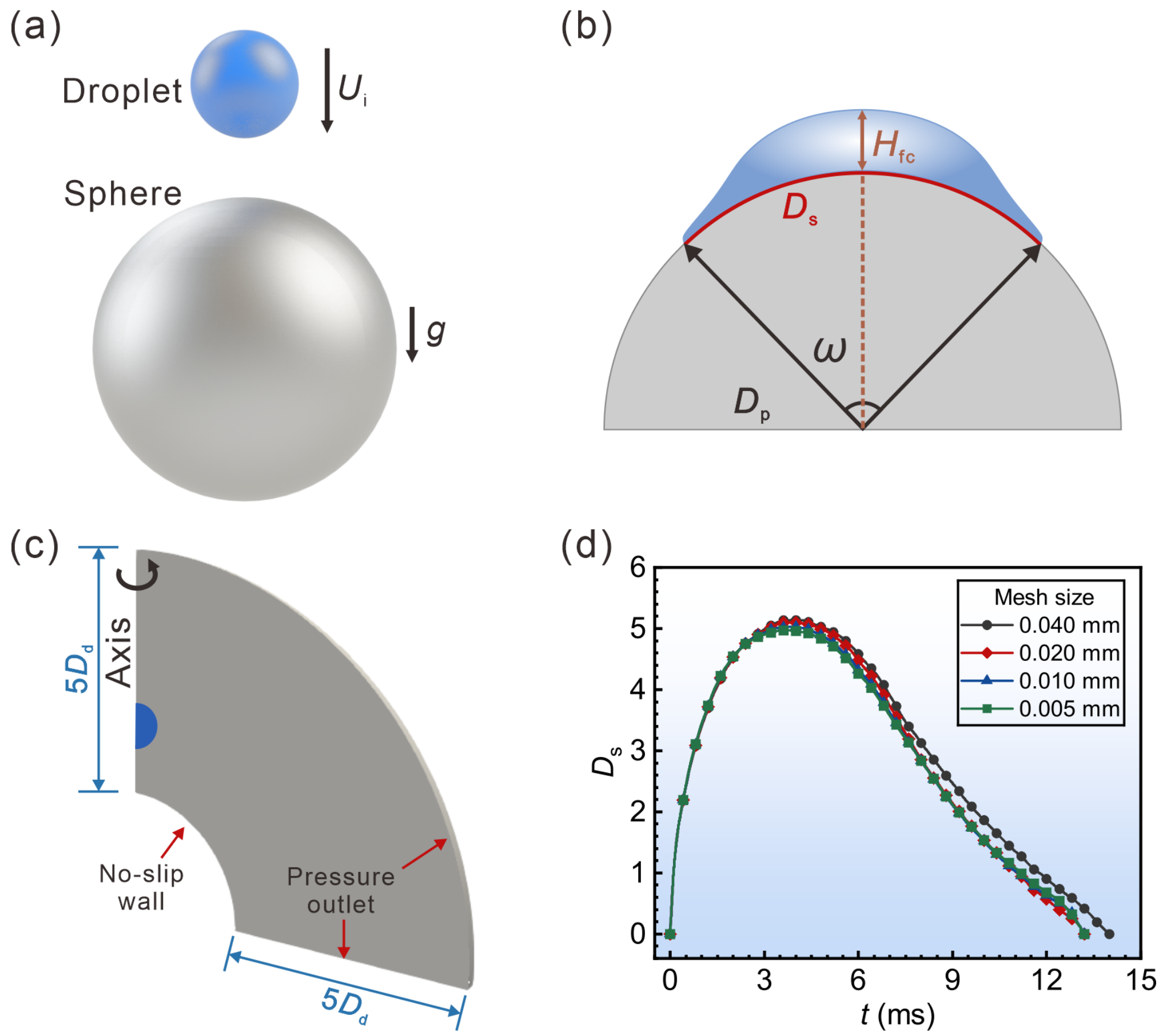

Figure 1. Physical model and mesh independence validation: (a) schematic of droplet impact on a sphere; (b) schematic of the geometric parameters during impact; (c) schematic of the computational domain; (d) evolution of the spreading length with time for different mesh resolutions. Here, $D_p$ = 10 mm, $D_d$ = 2.5 mm, $U_i$ = 1.0 m/s, $\mu_l$ = 0.898 mPa·s, $\mu_g$ = 0.0184 mPa·s, $\rho_l$ = 997 kg/m$^3$, $\rho_g$ = 1.184 kg/m$^3$, $\sigma$ = 0.072 N/m, and $\theta_{eq}$ =160°.

To assess the accuracy and reliability of the numerical model, the computational conditions are configured to match those employed in the reference experiments. The numerical predictions are then validated against the available experimental measurements. In terms of droplet morphology, the computed results show close agreement with the experimental observations reported by Ju et al. [65], as shown in Fig. 2(a). The spreading factor of a droplet impacting a spherical surface is quantitatively compared with several experimental results [22,65,66], as shown in Fig. 2(b). The simulated values are in good agreement with the measurements. These comparisons indicate that the present model can reliably reproduce droplet impact on a sphere.

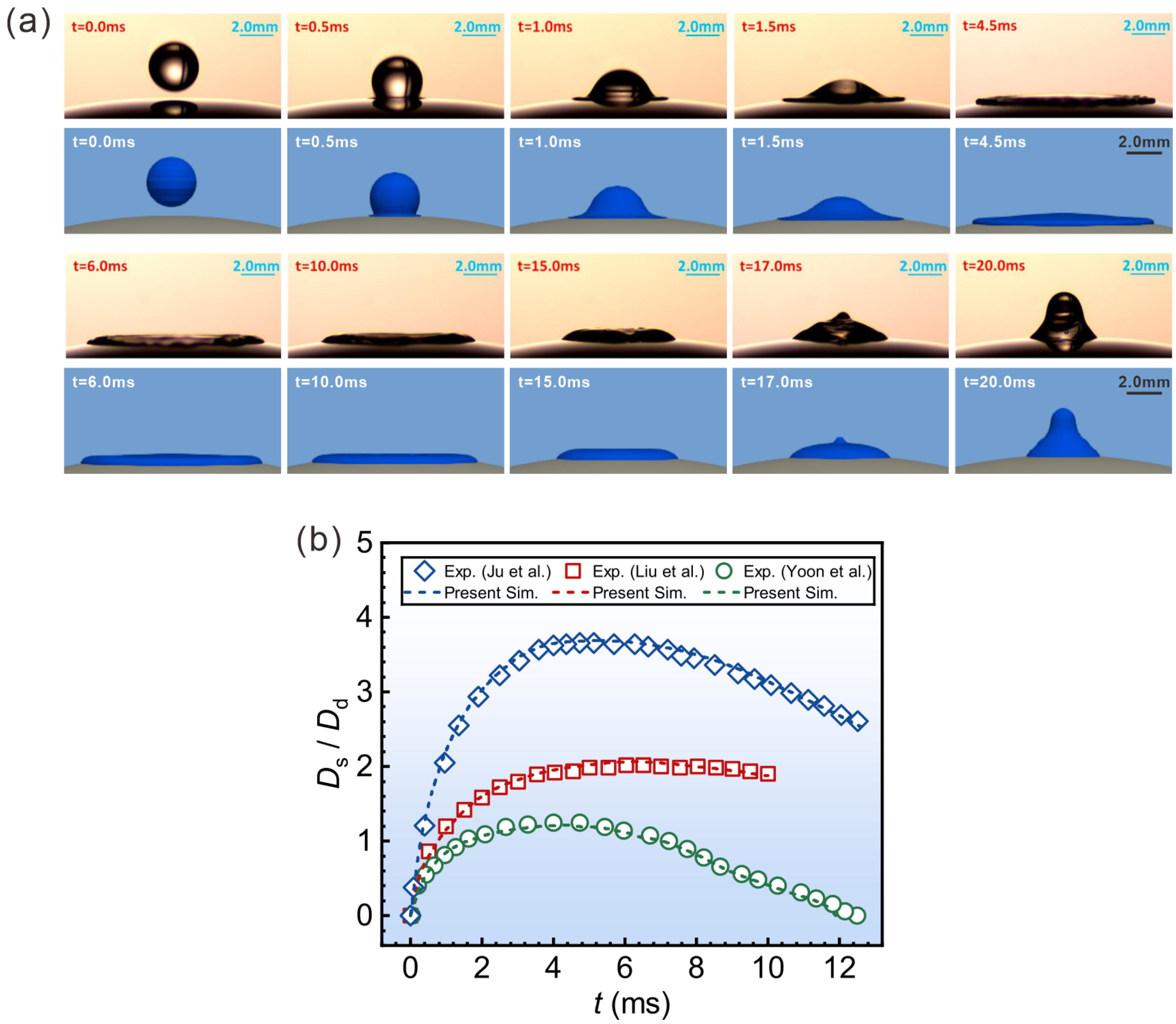

Figure 2. Comparison between simulations and experimental results [22,65,66]: (a) variation in droplet morphology (Reprinted with permission from [65] with permission from Elsevier); (b) variation in the dimensionless spreading length.

Unless otherwise stated, the following default parameters are used in the subsequent simulations. The gas density is $\rho_g$ = 1.184 kg/m$^3$, the liquid density is $\rho_l$ = 997 kg/m$^3$, the gas viscosity is $\mu_g$ = 0.0184 mPa·s, the surface tension coefficient is $\sigma$ = 0.072 N/m, and the droplet diameter is $D_d$ = 2 mm. The dimensionless parameters used in this work include the Weber number $We = \rho_l v_i^2 D_d / \sigma$, the Ohnesorge number $Oh = \mu_l / \sqrt{\rho_l D_d \sigma}$, and the dimensionless curvature $k^* = D_d / D_p$. In addition, to characterize the droplet morphology during impact, we defined the dimensionless spreading factor $D_s^* = D_s / D_d$ and the dimensionless central film thickness $H_{fc}^* = H_{fc} / D_d$. Here, $H_{fc}$ denotes the central film thickness measured along the axis of symmetry from the apex of the spherical surface to the droplet free surface, as presented in Fig. 1(b).

## 3. Results and discussion

### 3.1 Three typical impact modes of droplet impact on spherical surfaces

#### 3.1.1 Complete rebound mode

Three distinct impact outcomes are identified for droplets impinging on the spherical surface: complete rebound, partial rebound, and complete deposition. Three distinct impact regimes are identified according to the final attachment state of the liquid after retraction. Specifically, complete rebound is identified when the entire droplet detaches from the spherical surface and no residual liquid remains attached to the wall. Partial rebound is identified when only part of the liquid detaches from the surface, while a finite amount of liquid remains on the spherical surface. Complete deposition corresponds to the case in which no liquid detaches from the surface and the entire droplet remains attached after the impact and retraction process. Figure 3 illustrates the deformation of the droplet during the complete rebound mode. Immediately after contact, the bottom of the droplet is blocked by the wall. Its velocity decreases rapidly within a very short time, and a local high-pressure region forms in the contact area. However, the disturbance has not yet propagated to the upper part of the droplet. Therefore, the upper region still maintains its pre-impact motion. The velocity and pressure fields show no obvious change, as shown at 0.03 ms in Fig. 3(a). Afterwards, the droplet spreads along both sides of the spherical surface under the action of inertia. The high-velocity region moves toward the edge of the liquid film. The droplet height decreases rapidly, while the spreading length gradually increases. The vertical component of the droplet momentum decreases and eventually becomes zero. At 2.58 ms, the droplet reaches its maximum spreading state. At this moment, the pressure and velocity inside the droplet are relatively low, as shown at 2.58 ms in Fig. 3(a). Surface tension then drives the liquid film to retract. The liquid retracts and converges toward the central region, and the velocity and pressure near the droplet center increase, as shown at 5.18 ms in Fig. 3(a). Finally, the droplet forms an upwardly stretched liquid column and completely detaches from the wall at 9.03 ms.

To quantitatively describe the spreading process, Fig. 3(b) presents the time evolution of the dimensionless spreading factor $D_s^*$ and the dimensionless central film thickness $H_{fc}^*$. At the early stage of impact, the droplet spreads rapidly along the spherical surface due to inertia. The spreading factor increases quickly, while the central film thickness decreases significantly. Since the disturbance by the impact has not yet reached the upper part of the droplet, the top of the droplet still moves downward at nearly the initial impact velocity. As a result, the central film thickness shows an almost linear decrease at the early stage. After the droplet reaches the maximum spreading state, the leading edge retracts under the action of surface tension, and the spreading factor decreases. However, the central film thickness does not increase at the same time. Instead, the liquid in the central region is transferred outward under the combined effects of capillarity and the geometric constraint of the curved surface. Therefore, there is a clear time lag between the retraction of the spreading front and the re-accumulation of liquid in the central region.

Droplet impact on a spherical surface involves the continuous redistribution and interconversion of different forms of energy. These energies mainly include kinetic energy $E_{\mathrm{k}}$, gravitational potential energy $E_{\mathrm{g}}$, surface energy $E_{\sigma}$, and viscous dissipation $E_{\mathrm{diss}}$. To compare the relative contribution of each energy component during impact, the center of the sphere is taken as the reference level for gravitational potential energy. All energy components are normalized by the initial total energy of the droplet $E_{\mathrm{tot\text{-}0}}$. Figure 3(c) shows the temporal evolution of the normalized energies to reveal the energy conversion mechanism behind the droplet deformation. At the early stage of impact, most of the system energy exists in the form of kinetic energy $E_{\mathrm{k}}$. As the droplet contacts the spherical surface and spreads rapidly, the kinetic energy decreases markedly. Meanwhile, the surface energy $E_{\sigma}$ increases quickly, indicating that a large portion of the initial kinetic energy is converted into the surface energy required to create new interfaces. At the same time, the viscous dissipation $E_{\mathrm{diss}}$ continues to increase, showing that irreversible energy loss occurs during both spreading and retraction. Near the maximum spreading state, the kinetic energy reaches a low level, while the surface energy accounts for a relatively high proportion. The droplet is then strongly deformed. During the subsequent retraction process, part of the surface energy is converted back into kinetic energy. This drives the liquid to move toward the center and rebound from the surface. However, because viscous dissipation accumulates continuously, the energy stored at the interface cannot be fully recovered as kinetic energy. Over the whole impact process, the dissipated energy accounts for about 32%. In addition, the viscous dissipation increases rapidly during the spreading stage and the early retraction stage. When the droplet starts to stretch upward, the increase in viscous dissipation becomes slower. In contrast, the gravitational potential energy $E_{\mathrm{g}}$ changes only slightly during the whole process. This indicates that, over the present timescale, the droplet dynamics are mainly governed by inertia, surface tension, and viscous dissipation.

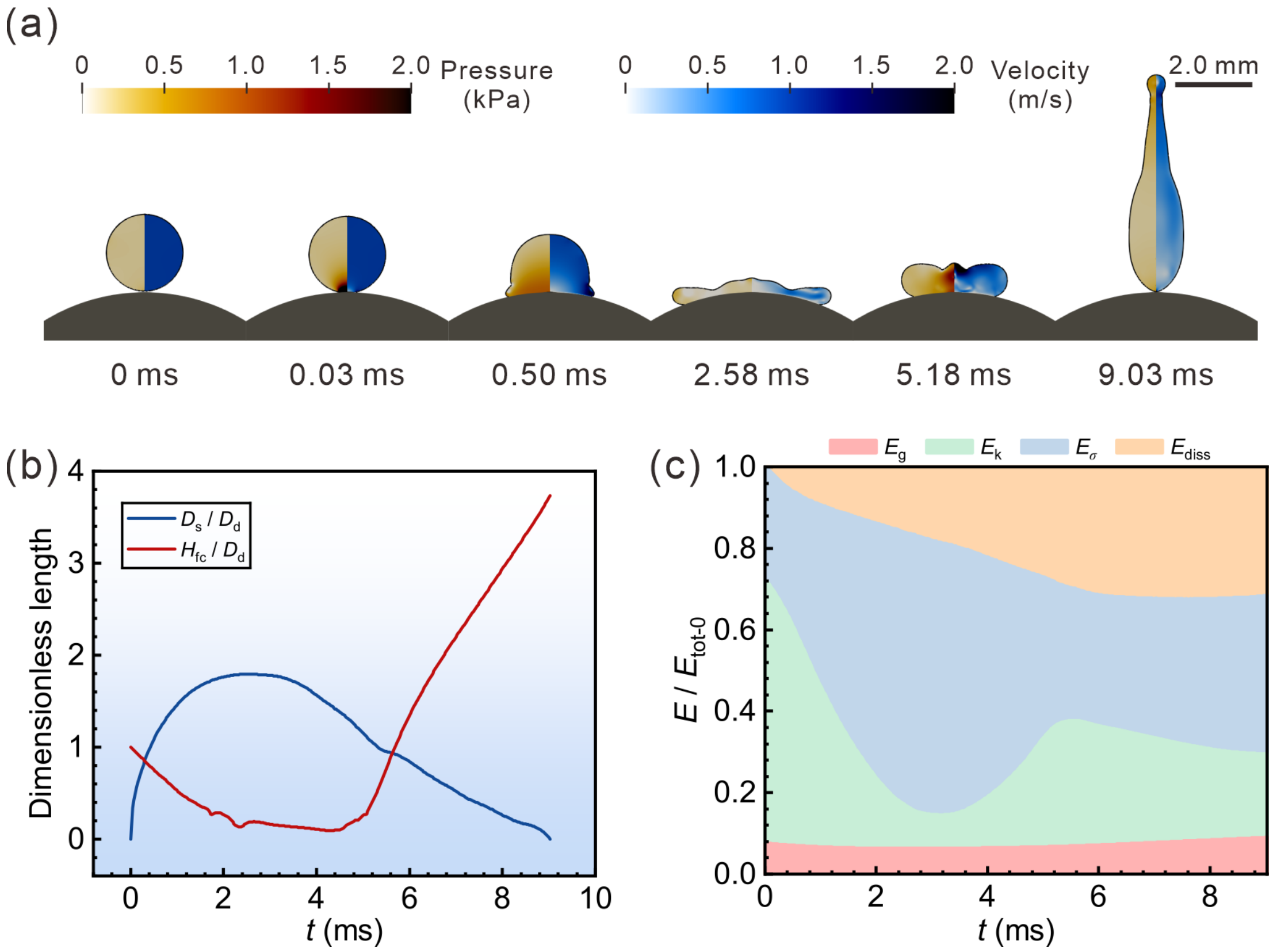

Figure 3. Complete rebound mode: (a) temporal evolution of droplet morphology and the associated flow fields, with velocity and pressure displayed on the right and left halves, respectively; (b) temporal variations in the spreading factor and central film thickness; (c) energy evolution during droplet impact. Here, $\theta_{eq}$ = 165°, $We$ = 27.7, $Oh$ = 0.00237, $k^*$ = 0.2.

To further examine the momentum transfer between the droplet and the spherical surface, as well as the wall force response in the complete rebound mode, Fig. 4 presents the temporal variations in droplet momentum, wall impact force, and its components. It also shows the spatiotemporal distributions of local pressure and shear stress along the arc-length direction. The wall impact force $F_{tot}$ is obtained indirectly from the temporal rate of change of droplet momentum. According to the force source, $F_{tot}$ can be decomposed into three parts: the pressure contribution $F_p$, the viscous contribution $F_v$, and the capillary contribution $F_{cl}$. As shown in Fig. 4(a), after the droplet contacts the spherical surface, its momentum decreases rapidly. The wall impact force rises sharply within a very short time and reaches the first peak at 0.50 ms, with $F_{t1}$ = 2.71 mN. This peak corresponds to the initial inertial impact stage. At this moment, the fluid at the bottom of the droplet is rapidly blocked by the spherical surface. The vertical momentum of the droplet decreases rapidly in the contact region and is transferred to the wall, resulting in a sharp increase in the wall impact force. As the vertical motion of the liquid changes into arcwise spreading along the spherical surface, the droplet momentum continues to decrease and reaches zero at 2.65 ms. The droplet then enters the retraction stage. Later, a second peak of $F_{tot}$ appears at 5.18 ms, with $F_{t2}$ =1.34 mN. As shown in Fig. 4(b), the variation in $F_p$ is generally consistent with that of $F_{tot}$ in the complete rebound mode. The contributions of $F_v$ and $F_{cl}$ are very small. Thus, the force acting on the wall is mainly contributed by the pressure force $F_p$.

Figures 4(c) and 4(d) show the spatiotemporal variations in local pressure and local shear stress along the spherical arc length, respectively. The local wall shear stress shown here represents the magnitude of the tangential viscous traction and is therefore always non-negative. In these plots, the horizontal axis represents the time after impact $t$, while the vertical axis $L_{arc}$ represents the spatial coordinate measured outward from the impact center along the spherical surface. The color indicates the magnitude of the local pressure or shear stress at a given time and position. The white curves represent the contact-line positions at different times, and the region below the white curves corresponds to the solid-liquid contact area. It can be seen that, during the initial contact stage, almost the entire solid-liquid contact region is a high-pressure region. It should be noted that the instantaneous maximum pressure may occur near the contact line immediately after contact, whereas the present description refers to the subsequent inertial impact stage, during which elevated pressure extends over much of the solid-liquid contact region. This stage corresponds to the formation of the first peak $F_{t1}$. As the spreading proceeds, the local pressure gradually decreases and propagates with the curved spreading front. At this stage, relatively high local pressure appears near both the spreading front and the central region. When the spreading length of the droplet reaches the droplet diameter, the local pressure decreases markedly. During retraction, the pressure distribution moves back toward the central region. In the liquid convergence process, a local high-pressure region forms again near the center, which corresponds to the occurrence of the second peak $F_{t2}$. Over the whole impact-rebound process, the central region experiences two local high-pressure events. They correspond to the initial inertial impact stage and the retraction convergence stage, respectively. Compared with the pressure field, the local shear stress is about one order of magnitude lower. At the early stage of impact, relatively high shear stress mainly appears near the contact line. This is because the velocity gradient of the near-wall fluid along the spherical surface first forms near the spreading front. As spreading develops, the spreading velocity gradually decreases. The near-wall velocity gradient becomes weaker, and the local shear stress also decreases. After the droplet enters the retraction stage, the shear stress shows a local increase again near the central region. This indicates that the liquid backflow and central convergence process re-induce a near-wall velocity gradient.

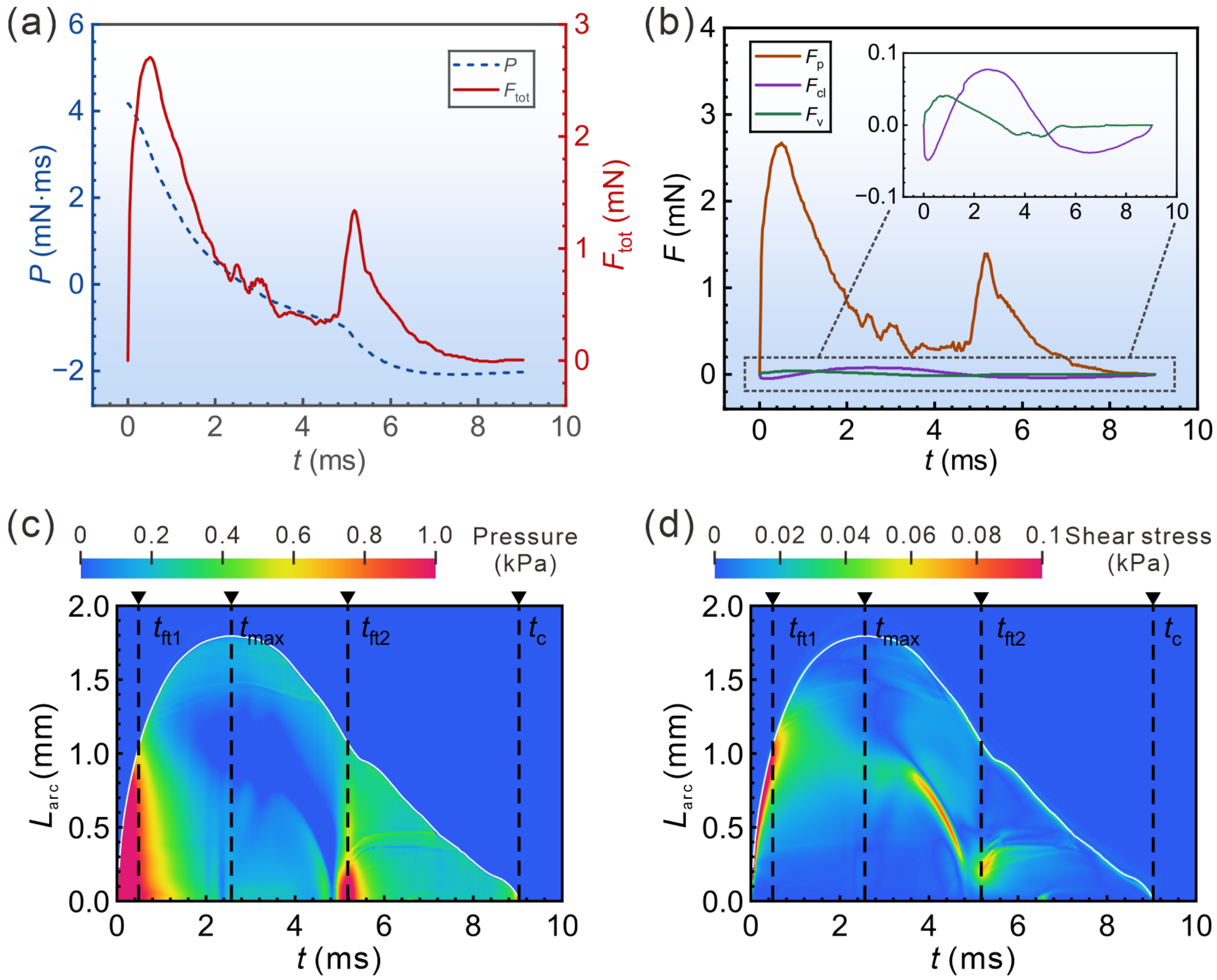

Figure 4. Transient force exerted on the wall during complete rebound: (a) droplet momentum and wall impact force during impact; (b) pressure force, capillary force, and viscous force acting on the wall during impact; (c) spatiotemporal evolution of local wall pressure along the arc-length direction; (d) spatiotemporal evolution of local wall shear stress along the arc-length direction. Here, $\theta_{eq} = 165°$, $We = 27.7$, $Oh = 0.00237$, $k^* = 0.2$.

### 3.1.2 Partial rebound mode

As the contact angle of the spherical surface is reduced to 75°, the droplet exhibits a partial rebound mode after impact, as shown in Fig. 5(a). Compared with the complete rebound mode, the stronger wettability of the wall enhances the adhesion between the droplet and the spherical surface. The droplet cannot completely detach from the wall during retraction. Instead, only part of the liquid breaks away during upward stretching, while the remaining liquid stays on the spherical surface. At the early stage of impact, the velocity and pressure fields are similar to those in the complete rebound mode. The main difference is that, because of the smaller contact angle, more gas is trapped beneath the droplet during contact. An air bubble forms near the wall, with a diameter of approximately 0.1 mm [67-69]. Afterwards, the bubble detaches from the hydrophilic wall, moves upward, and finally escapes. As shown in Figs. 5(b) and 5(d), the escape of the bubble causes oscillations in the central film thickness. At 4.50 ms, the spreading length reaches its maximum value. At this moment, both the pressure and velocity inside the droplet remain relatively low. During retraction, the droplet also forms a liquid column. However, compared with the complete rebound case, the liquid column becomes shorter. Finally, at 20.13 ms, breakup occurs near the top of the liquid column. The droplet separates into two parts. One part detaches from the sphere and forms a daughter droplet, while the remaining liquid stays attached to the spherical surface. Compared with the complete rebound mode, viscous dissipation increases more significantly in this mode and accounts for a larger fraction at the late stage, approximately 57%, as shown in Fig. 5(c). During retraction, part of the surface energy is still converted back into kinetic energy and drives the central liquid upward. However, less interfacial energy is available for conversion back into rebound kinetic energy, while viscous dissipation is enhanced. Therefore, the droplet cannot rebound as a whole and finally shows partial rebound.

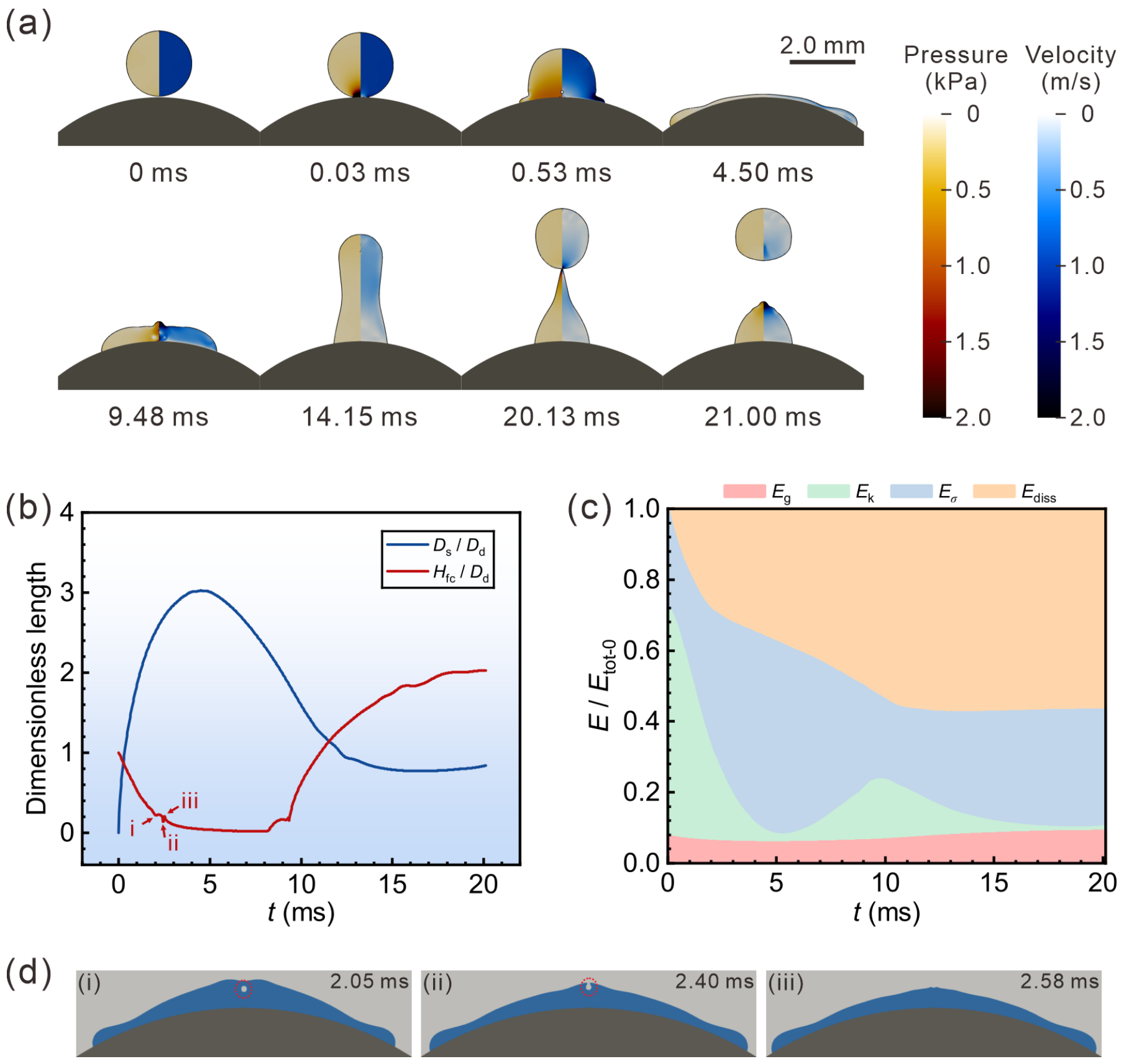

Figure 5. Partial rebound mode: (a) temporal evolution of droplet morphology and the associated flow fields, with velocity and pressure displayed on the right and left halves, respectively; (b) temporal variations in the spreading factor and central film thickness; (c) energy evolution during droplet impact; (d) bubble escape process corresponding to the three instants marked in (b). Here, $\theta_{eq} = 75°$, $We = 27.7$, $Oh = 0.00237$, $k^* = 0.2$.

The wall impact force $F_{tot}$ in the partial rebound mode still shows two peaks, as shown in Fig. 6(a). The first peak appears at 0.53 ms, with $F_{t1} = 2.51$ mN. The second peak appears at 9.48 ms, with $F_{t2} = 0.67$ mN. Compared with the complete rebound mode, the initial impact peak changes only slightly. This indicates that the momentum transfer during the initial inertial impact stage is similar in the two modes. In contrast, the second peak is much smaller than that in the complete rebound mode. This suggests that the secondary momentum exchange caused by liquid retraction and central convergence is significantly weakened in the partial rebound mode. Figure 6(b) shows that the viscous contribution $F_v$ remains small and becomes zero during the upward stretching stage of the liquid column. The capillary contribution $F_{cl}$ begins to increase, and its peak value is relatively large. It is more pronounced during spreading and in the early stage of retraction. The spatiotemporal evolutions of local pressure and local shear stress are similar to those in the complete rebound mode, as shown in Figs. 6(c) and 6(d). During the initial impact stage, the high-pressure region is mainly concentrated near the impact center, corresponding to the formation of the first peak. As the droplet spreads along the spherical surface, the pressure disturbance propagates outward with the spreading front and gradually weakens. When the droplet enters the retraction stage and starts to stretch upward, the pressure in the central region increases again. This corresponds to the formation of the second peak. The local shear stress is mainly concentrated near the contact line. As the spreading velocity decreases, the shear stress gradually weakens. During the upward stretching stage, both the pressure and shear stress become more uniformly distributed on the wall, and their magnitudes are relatively small.

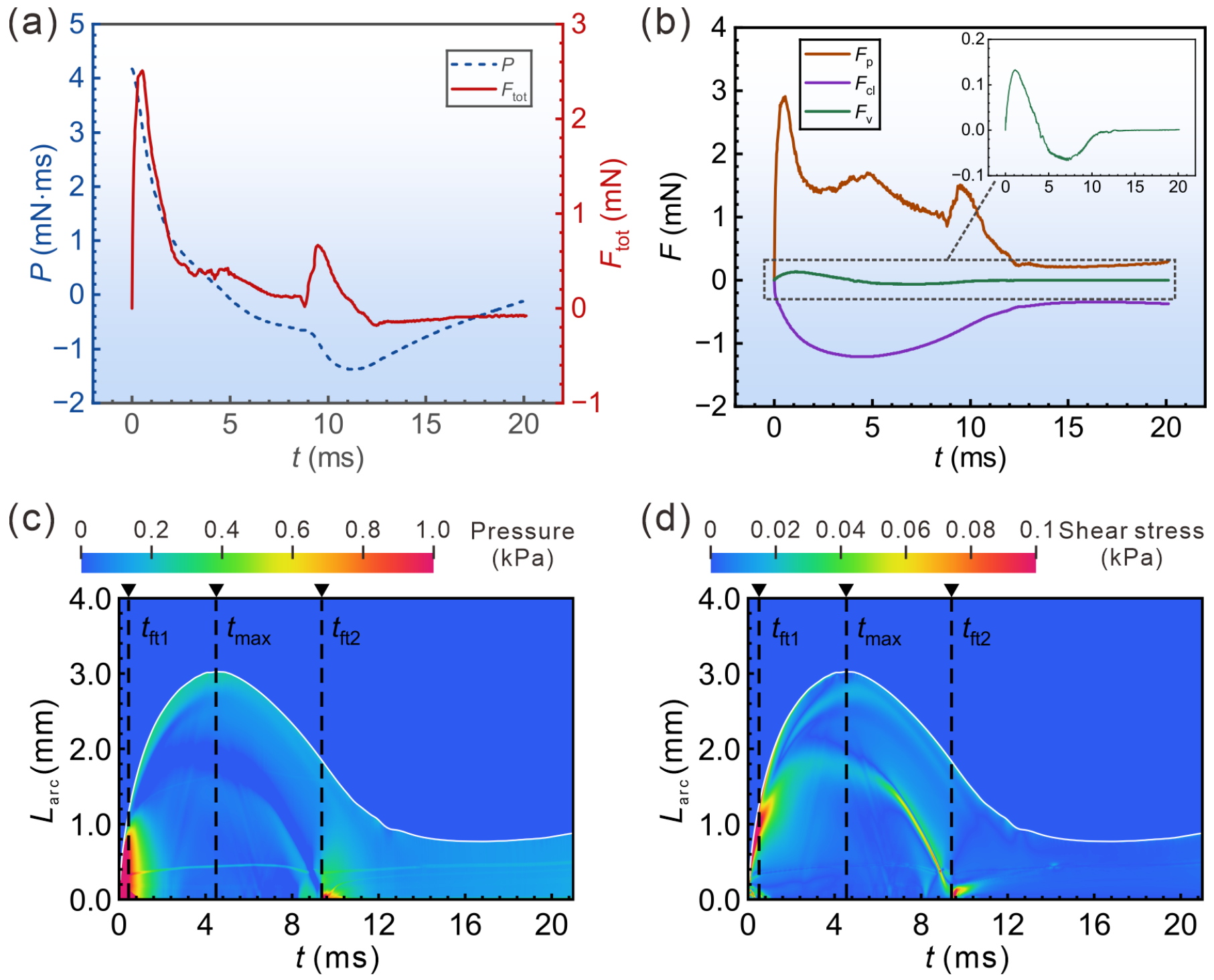

Figure 6. Transient force exerted on the wall during partial rebound: (a) droplet momentum and wall impact force during impact; (b) pressure force, capillary force, and viscous force acting on the wall during impact; (c) spatiotemporal evolution of local wall pressure along the arc-length direction; (d) spatiotemporal evolution of local wall shear stress along the arc-length direction. Here, $\theta_{eq} = 75°$, $We = 27.7$, $Oh = 0.00237$, $k^* = 0.2$.

### 3.1.3 Complete deposition mode

When the contact angle is further reduced to 45°, the droplet shows a complete deposition mode after impacting the spherical surface. Different from the complete rebound and partial rebound modes, the wall wettability is much stronger in this case. Although the droplet can still spread rapidly along the spherical surface under inertia at the early stage of impact, the subsequent retraction process is strongly suppressed. The liquid finally cannot detach from the sphere and remains deposited on the surface. At the early stage of impact, an air bubble with a diameter of approximately 0.1 mm still forms at the bottom of the droplet. The bubble then gradually detaches from the solid-liquid interface and escapes outward, causing fluctuations in the central film thickness, as shown in Figs. 7(b) and 7(d). As shown in Fig. 7(a), this mode produces the largest spreading length among the three impact outcomes, with its maximum attained at 6.28 ms. At the same time, the central film thickness decreases rapidly and remains at a relatively low level for a long period. A thin liquid film forms and persists in the central region. Finally, the droplet exhibits clear oscillations and then gradually becomes stable. The energy evolution in the complete deposition mode is shown in Fig. 7(c). The kinetic energy decays rapidly, while the surface energy and viscous dissipation increase quickly. In comparison with the two rebound modes, energy dissipation is the most pronounced in the complete deposition mode. The final dissipated energy accounts for about 80% of the initial total energy. Most of the energy is dissipated during the first spreading-retraction process. Although energy loss still occurs in the later oscillation stage, its magnitude is relatively small.

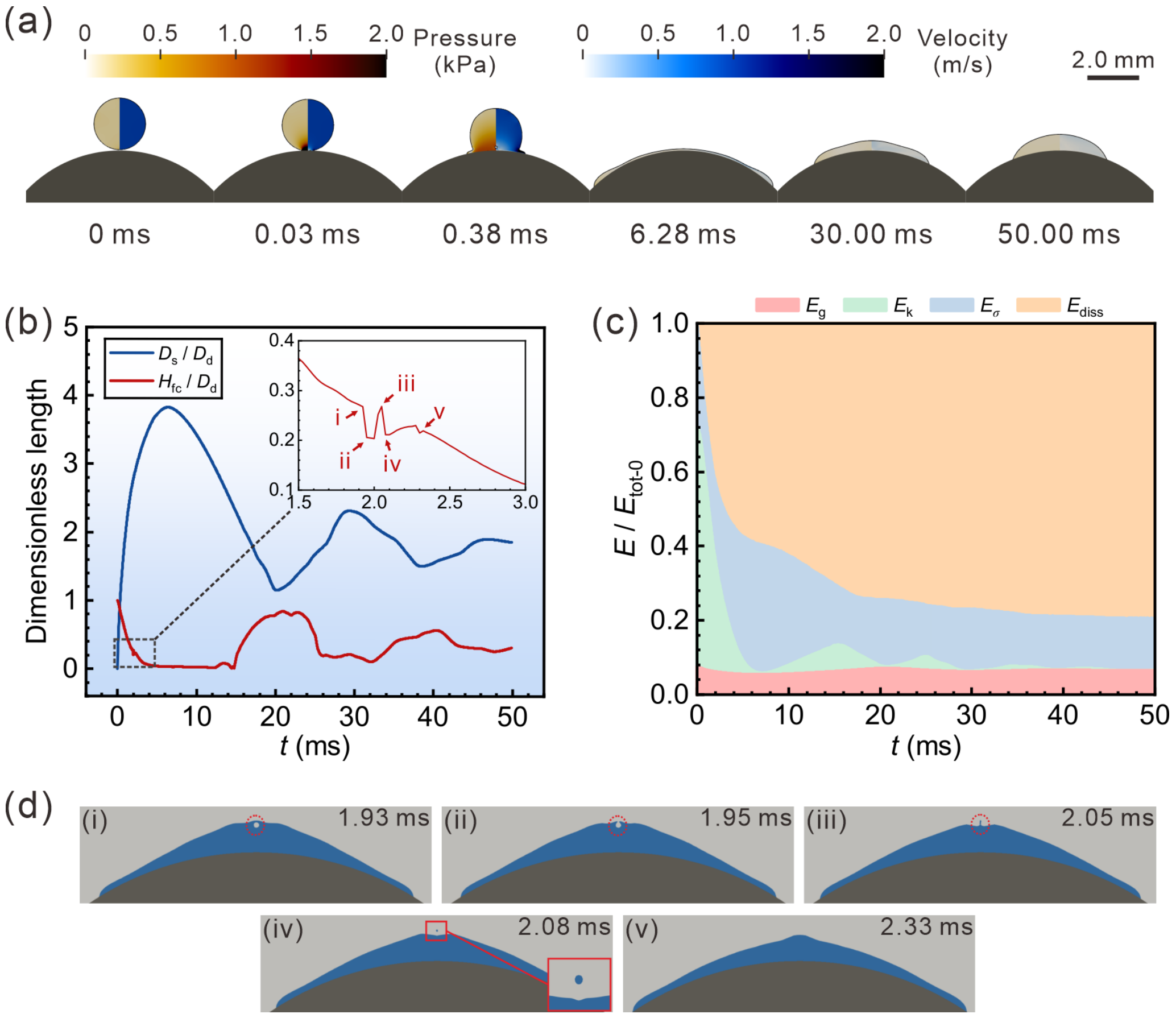

Figure 7. Complete deposition mode: (a) temporal evolution of droplet morphology and the associated flow fields, with velocity and pressure displayed on the right and left halves, respectively; (b) temporal variations in the spreading factor and central film thickness; (c) energy evolution during droplet impact; (d) bubble escape process corresponding to the five instants marked in (b). Here, $\theta_{eq} = 45°$, $We = 27.7$, $Oh = 0.00237$, $k^* = 0.2$.

Figure 8(a) shows the variation in the force acting on the spherical surface in the complete deposition mode. In this mode, the wall impact force $F_{tot}$ reaches a peak of 2.44 mN at 0.38 ms. Only one clear peak of impact force appears during the whole impact process, and no obvious second peak is observed. During the initial inertial-impact stage, the impact force is mainly associated with the pressure contribution $F_p$. Subsequently, the capillary contribution $F_{cl}$ increases markedly and becomes comparable to $F_p$. Because these two contributions act in opposite directions over much of the spreading-retraction process, they partially cancel each other. In contrast, the viscous contribution $F_v$ remains relatively small, as shown in Fig. 8(b). During the oscillation stage, the liquid film repeatedly spreads and retracts on the spherical surface. As a result, $F_{tot}$ and its components also show corresponding oscillatory features. The spatiotemporal distributions in Figs. 8(c) and 8(d) show that, during the first spreading-retraction process, the local pressure and local shear stress are similar to those in the complete rebound and partial rebound modes. A relatively high local pressure appears in the central region in the initial moments following impact. A relatively high local shear stress appears near the contact line. When the droplet retracts and the liquid column stretches upward, the pressure and shear stress in the central region are much lower than those in the previous two modes. Afterwards, the droplet enters the oscillation stage, and the distributions of pressure and shear stress become more uniform.

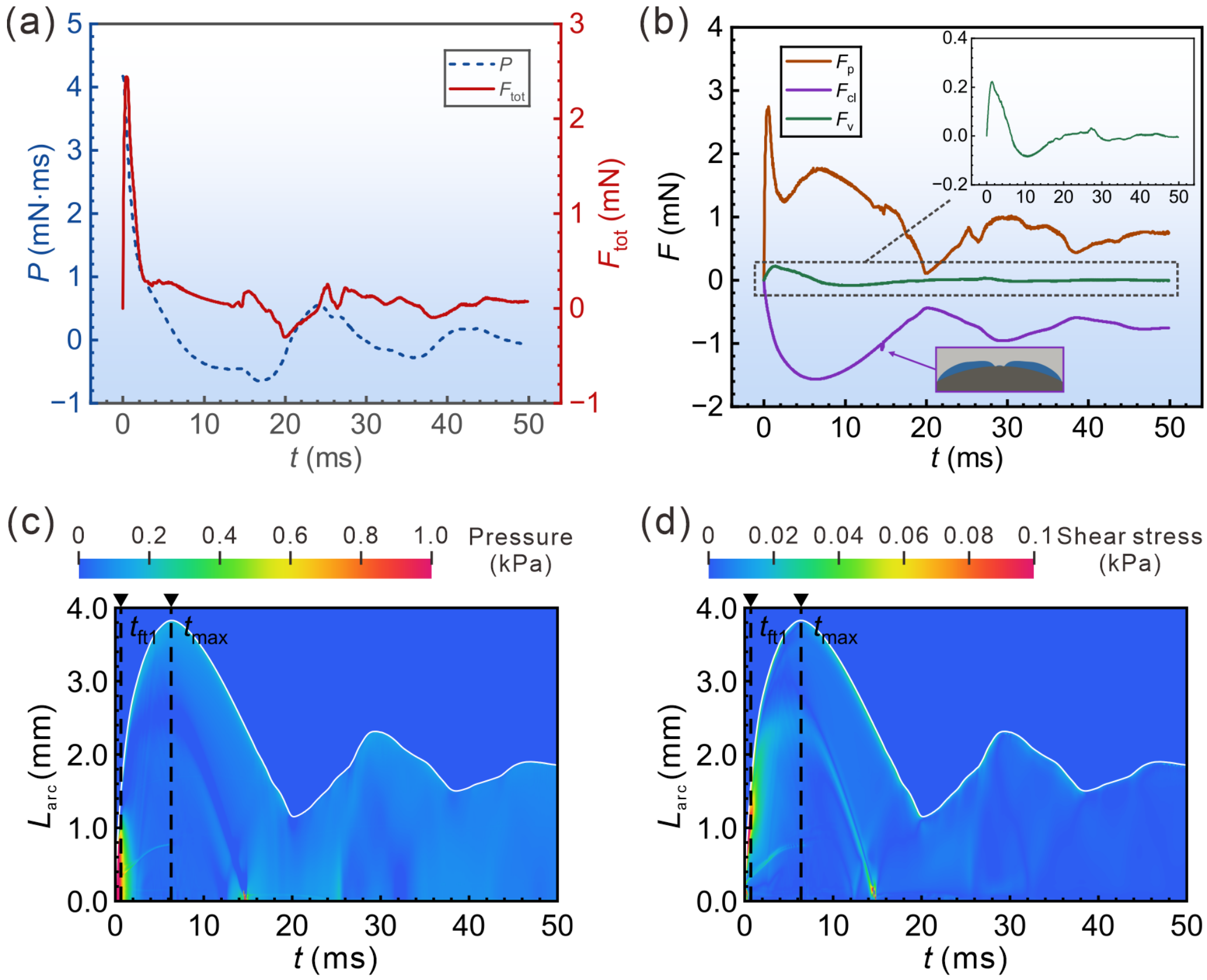

Figure 8. Transient force exerted on the wall during complete deposition: (a) droplet momentum and wall impact force during impact; (b) pressure force, capillary force, and viscous force acting on the wall during impact; (c) spatiotemporal evolution of local wall pressure along the arc-length direction; (d) spatiotemporal evolution of local wall shear stress along the arc-length direction. Here, $\theta_{eq} = 45°$, $We = 27.7$, $Oh = 0.00237$, $k^* = 0.2$.

### 3.2 Effect of wall wettability

Wall wettability plays a crucial role in droplet impact behavior. It governs how the droplet spreads, retracts, and adheres to the surface. It also influences the post-impact liquid-film evolution and interfacial morphology. To clarify the role of wettability in droplet-sphere interactions, the equilibrium contact angle $\theta_{eq}$ is varied in this section. As discussed in Section 3.1, with increasing contact angle, i.e., decreasing wettability, the impact mode changes from complete deposition to partial rebound and then to complete rebound. When a droplet impacts a spherical surface, the maximum spreading length $D_{s\text{-}max}$ and the time to reach maximum spreading $t_{max}$ are mainly governed by inertia and surface tension. Therefore, the droplet diameter $D_d$ and the inertial-capillary time $t_\sigma = \left(\rho_l R_d^3/\sigma\right)^{1/2}$ are used to normalize the length and the time, respectively. Here, $R_d = D_d / 2$ is the droplet radius. As wettability decreases, both $D_{s\text{-}max} / D_d$ and $t_{max} / t_\sigma$ decrease, as shown in Fig. 9(a). At the early stage of impact, droplet spreading is mainly driven by inertia. As the spreading length increases, the capillary-driven retraction near the contact line gradually becomes stronger.

When the droplet spreads over a one-sided arc length $R_s = D_s/2$ along the spherical surface, the generated surface energy can be written as

$$E_s = 2\pi\sigma R_p^2\left(1-\cos\left(R_s/R_p\right)\right)\left(1-\cos\theta_{eq}\right), \tag{29}$$

where $R_p = D_p / 2$ is the radius of the sphere. The capillary force acting on the liquid film is then given by

$$F_{cl} = \partial E_s/\partial R_s = 2\pi\sigma R_p \sin\left(R_s/R_p\right)\left(1-\cos\theta_{eq}\right). \tag{30}$$

Thus, under the same impact condition, an increase in the contact angle strengthens the braking effect of surface tension. This causes the contact line velocity to decrease to zero earlier. Therefore, a larger contact angle makes the droplet enter the retraction stage earlier from the spreading stage. The droplet reaches a smaller maximum spreading arc length in a shorter time. In addition, with increasing contact angle, the rate of change of $\left(1-\cos\theta_{eq}\right)$ gradually decreases. This indicates that the strengthening effect of further increasing the contact angle becomes limited. Therefore, the influence of contact angle on the time to reach maximum spreading $t_{max}$ becomes weaker, and $t_{max}$ enters a relatively stable range. Therefore, at relatively large contact angles, namely in the complete rebound regime, the time to maximum spreading is mainly governed by the initial inertia-capillary timescale

$$t_{max} \sim t_\sigma . \tag{31}$$

Next, the effect of wettability on the contact time in the complete rebound mode is considered. It should be noted that the numerical simulations account for the instantaneous dynamic contact angle through the modified Kistler model. However, in the following theoretical analysis, the equilibrium contact angle $\theta_{eq}$ is adopted as a characteristic wettability parameter to obtain a simplified scaling relation, rather than to describe the instantaneous contact line dynamics during retraction. When the droplet just enters the retraction stage, the inertial force $F_i$ is balanced by the capillary force $F_{cl}$

$$F_i = \frac{d}{dt}\left( m_f \frac{dr_s}{dt} \right) = F_{cl}, \tag{32}$$

where $r_s$ is the retraction distance over a short time, and $U_{ret} = dr_s/dt$ is the retraction velocity. Assuming that the liquid film thickness $h_f$ is approximately uniform, the liquid film mass $m_f$ can be written as

$$m_f = \rho_l h_f \left( 2\pi R_p \sin\left( R_s/R_p \right) \right) r_s. \tag{33}$$

Substituting Eq. (33) into Eq. (32) gives

$$\frac{d}{dt}\left( r_s \frac{dr_s}{dt} \right) = \frac{\sigma\left(1-\cos\theta_{eq}\right)}{\rho_l h_f}. \tag{34}$$

The retraction velocity can then be obtained as

$$U_{ret} = \left[ \frac{\sigma\left(1-\cos\theta_{eq}\right)}{\rho_l h_f} \right]^{1/2}. \tag{35}$$

Once the droplet attains its maximum spreading extent, the wetted spherical cap area $S_{cap}$ is

$$S_{cap} = \frac{\pi D_p^2}{2}\left[ 1-\cos\left( \frac{D_{s\text{-}max}}{D_p} \right) \right]. \tag{36}$$

At this moment, the liquid film thickness $h_f$ can be estimated from the initial droplet volume $V_d$ and the wetted spherical cap area $S_{cap}$

$$h_f \sim \frac{V_d}{S_{cap}} \sim \frac{D_d^3}{3D_p^2\left[ 1-\cos\left( \frac{D_{s\text{-}max}}{D_p} \right) \right]}. \tag{37}$$

Substituting Eq. (37) into Eq. (35), the retraction velocity $U_{ret}$ can be written as

$$U_{\mathrm{ret}} \sim D_{\mathrm{p}}\left[1-\cos\left(\frac{D_{\mathrm{s-max}}}{D_{\mathrm{p}}}\right)\right]^{1/2}\left(\frac{\rho_1 D_{\mathrm{d}}^3}{\sigma}\right)^{-1/2}\left(1-\cos\theta_{\mathrm{eq}}\right)^{1/2}. \tag{38}$$

Therefore, the retraction time $t_{\mathrm{ret}}$ can be estimated as

$$t_{\mathrm{ret}} \sim \frac{D_{\mathrm{s-max}}/2}{U_{\mathrm{ret}}} \sim \left(\frac{\rho_1 D_{\mathrm{d}}^3}{\sigma}\right)^{1/2}\frac{D_{\mathrm{s-max}}/D_{\mathrm{p}}}{\left[1-\cos\left(D_{\mathrm{s-max}}/D_{\mathrm{p}}\right)\right]^{1/2}}\left(1-\cos\theta_{\mathrm{eq}}\right)^{-1/2}. \tag{39}$$

In the complete rebound mode, $D_{\mathrm{s-max}}/D_{\mathrm{p}} \ll 1$ . Therefore, applying Taylor expansion to $\cos\left(D_{\mathrm{s-max}}/D_{\mathrm{p}}\right)$, we can obtain

$$\frac{D_{\mathrm{s-max}}/D_{\mathrm{p}}}{\left[1-\cos\left(D_{\mathrm{s-max}}/D_{\mathrm{p}}\right)\right]^{1/2}} \approx \frac{D_{\mathrm{s-max}}/D_{\mathrm{p}}}{\left[\frac{1}{2}\left(D_{\mathrm{s-max}}/D_{\mathrm{p}}\right)^2\right]^{1/2}} = \sqrt{2}. \tag{40}$$

Ultimately, we arrive at

$$t_{\mathrm{ret}}/t_\sigma \sim \left(1-\cos\theta_{\mathrm{eq}}\right)^{-1/2}. \tag{41}$$

During droplet impact, the contact time $t_{\mathrm{c}}$ consists of the spreading time $t_{\mathrm{max}}$ and the retraction time $t_{\mathrm{ret}}$. According to Eq. (31), $t_{\mathrm{max}}$ can be treated as nearly constant. In the complete rebound regime, the variation in $t_{\mathrm{max}}$ / $t_\sigma$ is relatively weak, while the dependence of $t_{\mathrm{c}}$ on wettability is mainly associated with the retraction stage. According to Eq. (41), the wettability-dependent contribution scales with $\left(1-\cos\theta_{\mathrm{eq}}\right)^{-1/2}$. Since the physically admissible contact angle satisfies $\theta_{\mathrm{eq}} \leq 180°$, the minimum value of $\left(1-\cos\theta_{\mathrm{eq}}\right)^{-1/2}$ is $1/\sqrt{2}$, corresponding to the limiting non-wetting state $\theta_{\mathrm{eq}} = 180°$. Taking this state as the reference, the contact time correlation can therefore be expressed as

$$t_{\mathrm{c}}/t_\sigma = C_1\left[\left(1-\cos\theta_{\mathrm{eq}}\right)^{-1/2} - \frac{1}{\sqrt{2}}\right] + C_2, \tag{42}$$

where $C_1$ and $C_2$ are constants and $C_2$ represents the dimensionless contact time in the limiting non-wetting state $\theta_{\mathrm{eq}} = 180°$. As shown in Fig. 9(b), the fitting curve of Eq. (42) $t_{\mathrm{c}}/t_\sigma = 7.45\left[\left(1-\cos\theta_{\mathrm{eq}}\right)^{-1/2} - 1/\sqrt{2}\right] + 2.34$, represented by the red dashed line, shows close agreement with the numerical predictions. This indicates that the proposed contact time model agrees well with the variation of the contact time with surface wettability in the complete rebound mode.

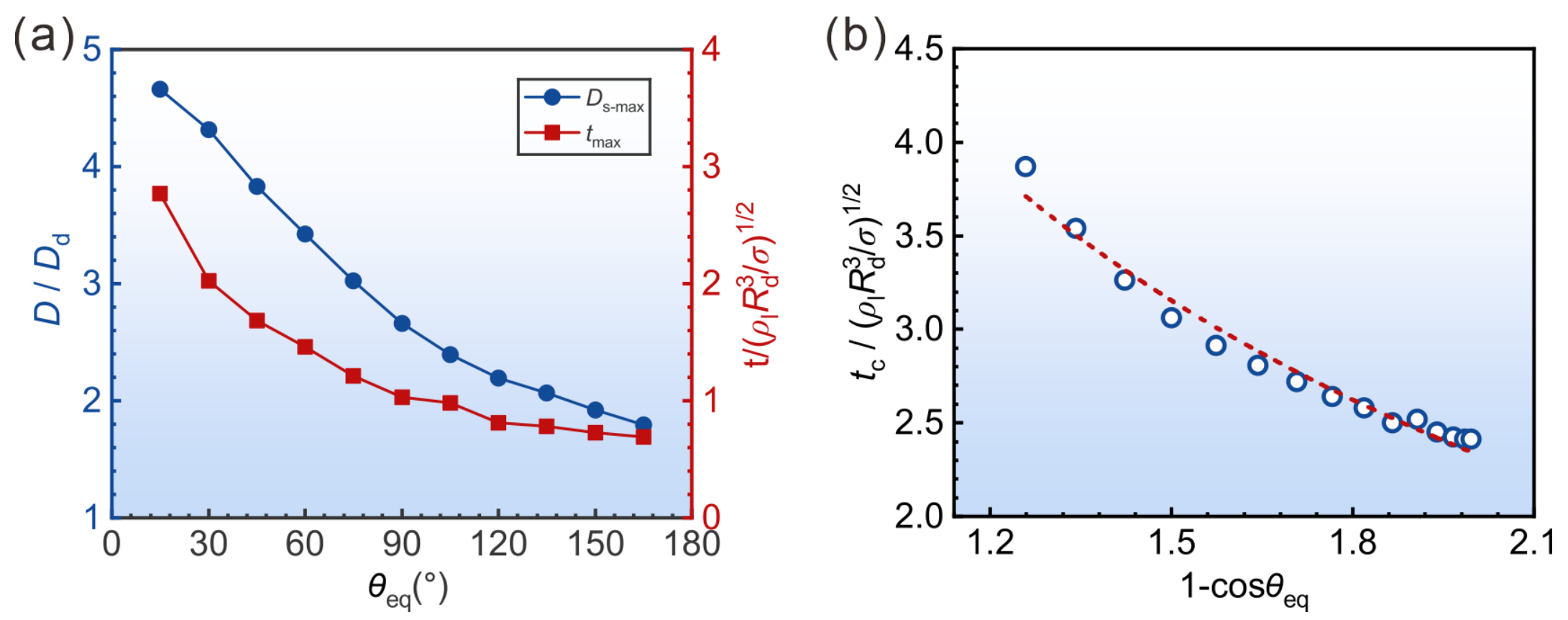

Figure 9. Effect of wall wettability on the kinematic parameters: (a) maximum spreading length $D_{s\text{-max}}$ and the corresponding time $t_{max}$; (b) contact time $t_c$ in the complete rebound mode. Here, the length and time are normalized by the droplet diameter $D_d$ and the inertial-capillary time $t_\sigma = \left(\rho_l R_d^3/\sigma\right)^{1/2}$, respectively. $We = 27.7$, $Oh = 0.00237$, $k^* = 0.2$.

As shown in Fig. 10(a), the first impact force peak $F_{t1}$ shows a relatively weak dependence on surface wettability, whereas its occurrence time $t_{ft1}$ exhibits a moderate variation over the investigated range. Specifically, $t_{ft1}$ changes from approximately 0.38 ms to 0.50 ms as the contact angle increases. Nevertheless, compared with the second peak characteristics, the first-peak parameters remain less sensitive to wettability. This is because the first force peak is mainly governed by the initial inertial momentum transfer. Immediately after initial contact, the droplet has not yet undergone obvious spreading, and the contact line moves over only a limited range. Therefore, the effect of surface tension is limited. The first impact force peak is mainly determined by the inertial effect at the initial moment of impact. With decreasing wettability, $F_{t1}$ and $t_{ft1}$ increase slightly. This is because stronger hydrophobicity suppresses droplet spreading on the wall to some extent. After normalization by $\rho_l U_i^2 D_d^2$, the dimensionless first peak $F_{t1}^*$ eventually approaches 0.68, and $t_{ft1}$ / $t_\sigma$ remains close to 0.13. Different from the first peak, the second impact force peak $F_{t2}$ is more sensitive to surface wettability, as shown in Fig. 10(b). This is because the second peak appears during the retraction stage, which is closely related to surface tension. Therefore, wall wettability affects the second peak of the impact force. In the complete rebound mode, $F_{t2}$ first increases with increasing contact angle and reaches its maximum value near 135°. It then decreases as the wettability is further reduced. Figure 10(c) further shows the variation in the occurrence time of the second peak $t_{ft2}$ with contact angle. Similar to the analysis of the contact time $t_c$ in Eq. (42), $t_{ft2}$ also consists of the spreading stage and the retraction stage

$$t_{ft2}/t_\sigma = C_3\left[\left(1-\cos\theta_{eq}\right)^{-1/2} - \frac{1}{\sqrt{2}}\right] + C_4 . \tag{43}$$

The fitted relation in Eq. (43) is $t_{\mathrm{ft2}}/t_\sigma = 2.35\left[\left(1-\cos\theta_{\mathrm{eq}}\right)^{-1/2} - 1/\sqrt{2}\right]+1.38$, plotted as a red dashed line in Fig. 10(c). This relation agrees well with the numerical results. It shows that surface wettability mainly affects the occurrence time of the second impact force peak by changing the retraction stage. The proposed model can therefore describe the variation in $t_{\mathrm{ft2}}$ with surface wettability.

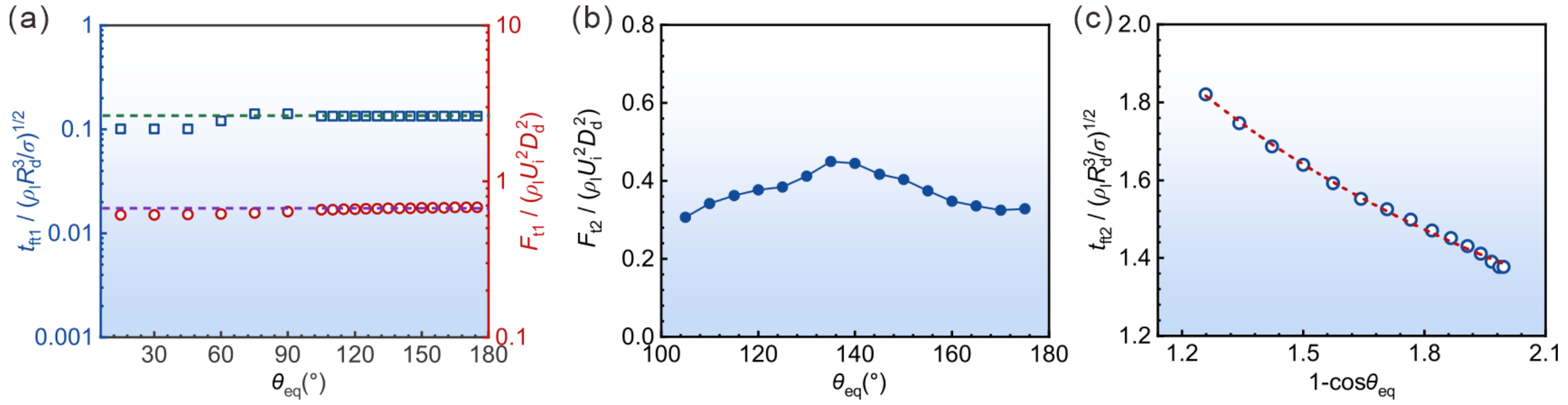

Figure 10. Effect of wall wettability on the dynamic parameters: (a) variation of the first impact force peak $F_{\mathrm{t1}}$ and its occurrence time $t_{\mathrm{ft1}}$ with wall wettability; (b) variation of the second impact force peak $F_{\mathrm{t2}}$ with wall wettability in the complete rebound mode; (c) variation of the occurrence time of the second peak $t_{\mathrm{ft2}}$ with wall wettability. Here, the time and force are normalized by the inertial-capillary time $t_\sigma = \left(\rho_l R_d^3/\sigma\right)^{1/2}$ and $\rho_l U_i^2 D_d^2$, respectively. $We = 27.7$, $Oh = 0.00237$, $k^* = 0.2$.

### 3.3 Effect of spherical curvature

For droplet impact on a sphere, the dimensionless curvature $k^* = D_d/D_p$ is an important geometric parameter for characterizing the curvature effect. In this study, $k^*$ is varied by changing the sphere diameter, with a variation range of $0 \sim 0.5$. The case of 0 corresponds to impact on a flat wall. As shown in Fig. 11(a), when $k^*$ is small, the curvature of the spherical surface in the contact region is weak. The impact process is therefore close to that on a locally flat surface. In this case, the maximum spreading length $D^*_{\text{s-max}}$ and the corresponding dimensionless time $t_{\max} / t_\sigma$ change only slightly. When $k^* > 0.06$, the curvature effect becomes stronger. Both $D^*_{\text{s-max}}$ and $t_{\max} / t_\sigma$ increase as the dimensionless curvature increases. The dependence of the spreading arc angle on dimensionless curvature is presented in Fig. 11(b). The relationship is given by $\omega_{\max}/2\pi \sim k^*$, with a coefficient of approximately 0.60. The dimensionless curvature has no obvious effect on the contact time, as shown in Fig. 11(c). The value of $t_c / t_\sigma$ remains close to 2.5. This indicates that the time-scale of droplet rebound is still mainly governed by inertial-capillary dynamics and is almost independent of the size of the sphere.

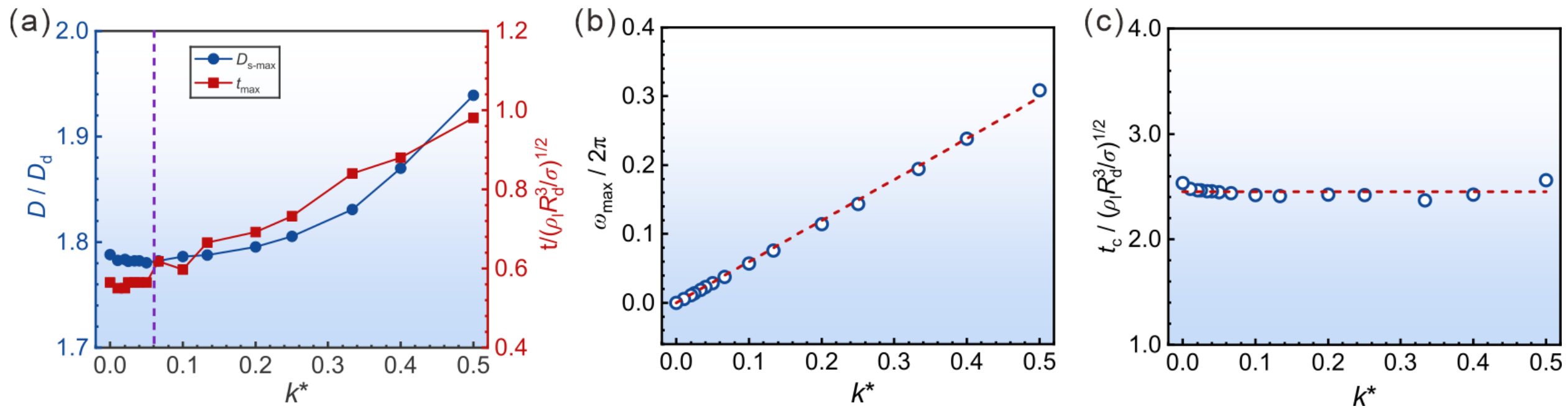

Figure 11. Effect of the dimensionless curvature $k^*$ on the kinematic parameters: (a) maximum spreading length $D_{s\text{-max}}$ and corresponding time $t_{max}$; (b) maximum spreading angle $\omega_{max}$; (c) contact time $t_c$. Here, the length, angle, and time are normalized by the droplet diameter $D_d$, $2\pi$, and the inertial-capillary time $t_\sigma = \left(\rho_l R_d^3/\sigma\right)^{1/2}$, respectively. $\theta_{eq}$ = 165°, $We$ = 27.7, $Oh$ = 0.00237.

The effect of the dimensionless curvature $k^*$ on the impact force is shown in Fig. 12. The first impact force peak $F_{t1}$ decreases significantly as $k^*$ increases, as shown in Fig. 12(a). When the spherical curvature is small, the normal momentum of the droplet can be transferred to the solid surface within a short time. This produces a relatively large initial impact force peak. As the spherical curvature increases, the droplet tends to spread tangentially along the spherical surface after contact. Part of the incident momentum is converted into liquid-film motion along the surface. This weakens the transfer of normal momentum to the sphere. Previous studies on droplet impact on flat surfaces have shown that $F_{t1} \sim C_3 \rho_l U_i^2 D_d^2$ [36,39], where $C_3 \approx 0.84$ for the flat wall case in the present study. For a spherical surface, a curvature-related coefficient is introduced as $C_{f1} = C_{f1}\left(k^*\right)$, and $C_{f1} \to 1$ when $k^* \to 0$. By applying a Taylor expansion to $C_{f1}$, one obtains $C_f = 1 + C_{f1}\left(k^*\right) + O\left[\left(k^*\right)^2\right]$. Thus, the dimensionless first impact force peak $F_{t1}^*$ can be approximated as

$$F_{t1}^* = \frac{F_{t1}}{\rho_l U_i^2 D_d^2} \sim C_3\left[1 + C_{f1}\left(k^*\right)\right], \tag{44}$$

Fitting the simulation results gives $C_{f1}$ = -0.87. Therefore,

$$F_{t1}^* = 0.84\left[1 - 0.87\left(k^*\right)\right] = 0.84 - 0.73k^*. \tag{45}$$

Equation (45) shows that the dimensionless first impact force peak $F_{t1}^*$ is linearly and negatively correlated with the dimensionless curvature $k^*$. When $k^*$ is large, the spherical curvature is strong. The transfer of normal momentum is weakened, leading to a decrease in $F_{t1}^*$. As $k^*$ decreases, the curvature effect gradually weakens, and $F_{t1}^*$ increases progressively, approaching the flat wall impact limit of 0.84. Different from $F_{t1}^*$, the dimensionless second impact force peak $F_{t2}^* = F_{t2} / \rho_l U_i^2 D_d^2$ does not vary monotonically. As the dimensionless curvature increases, $F_{t2}^*$ first decreases and then increases, reaching a relatively low value near the intermediate dimensionless curvature of $k^* = 0.2$, as shown in Fig. 12(c). The dimensionless curvature $k^*$ has only a weak effect on the occurrence time of the first peak $t_{ft1} / t_\sigma$. It only shows slight fluctuations and remains close to 0.13, as illustrated in Fig. 12(b). In contrast, the occurrence time of the second peak $t_{ft2} / t_\sigma$ shows a linear relationship with $k^*$, which can be expressed as:

$$t_{ft2}/t_\sigma = 0.94k^* + 1.2\,. \tag{46}$$

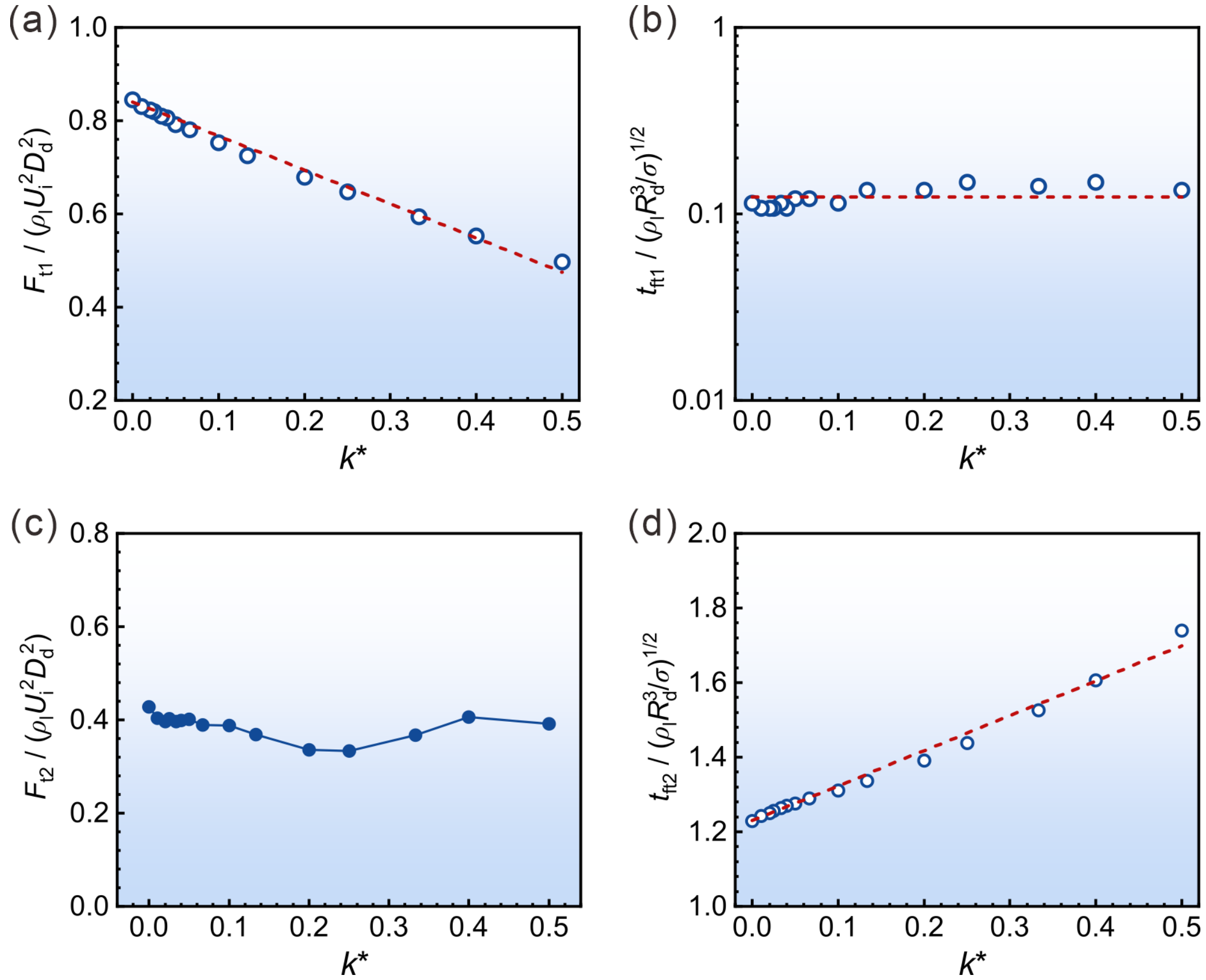

Figure 12. Effect of the dimensionless curvature $k^*$ on the dynamic parameters: (a) first impact force peak $F_{t1}$; (b) occurrence time of the first peak $t_{ft1}$; (c) second impact force peak $F_{t2}$; (d) occurrence time of the second peak $t_{ft2}$. Here, the force and time are normalized by $\rho_l U_i^2 D_d^2$ and the inertial-capillary time $t_\sigma = \left(\rho_l R_d^3/\sigma\right)^{1/2}$, respectively. $\theta_{eq} = 165°$, $We = 27.7$, $Oh = 0.00237$.

Figure 13 shows the variation in the critical contact angle for complete rebound with the dimensionless curvature. The critical contact angle is defined here as the minimum equilibrium contact angle required for complete rebound. For each dimensionless curvature, the critical contact angle was determined using a bisection method. Two initial contact angles were selected to bracket the transition between complete rebound and non-complete-rebound states. The midpoint of the interval was then simulated, and the interval was updated according to the resulting impact mode. This procedure was repeated until the interval width was reduced to 0.1°. For a flat wall, the critical contact angle is 105.3°. When the sphere diameter decreases to 6 mm, the critical contact angle decreases to 94.5°. Within the investigated range $0 \le k^* \le 0.33$, the contact time does not change with surface curvature, the critical contact angle for complete rebound decreases as the spherical curvature increases. Whether the droplet can completely rebound depends on whether the surface energy stored and released during retraction can overcome solid-liquid adhesion and viscous dissipation. The adhesion work per unit solid-liquid area can be approximated as $W_a = \sigma\left(1+\cos\theta_{eq}\right)$ [70]. Therefore, a smaller contact angle corresponds to stronger solid-liquid adhesion, making droplet detachment more difficult. For a flat wall, the droplet must directly overcome a strong normal adhesion resistance during retraction and detachment. Complete rebound can occur only when the contact angle is sufficiently large. For a spherical surface, however, as the curvature increases, the solid-liquid contact region after spreading is distributed over a curved surface with different local normal directions. The effective component of adhesion in the overall detachment direction is therefore weakened. As a result, increasing spherical curvature helps the droplet detach from the wall at a relatively low contact angle, thereby promoting complete rebound.

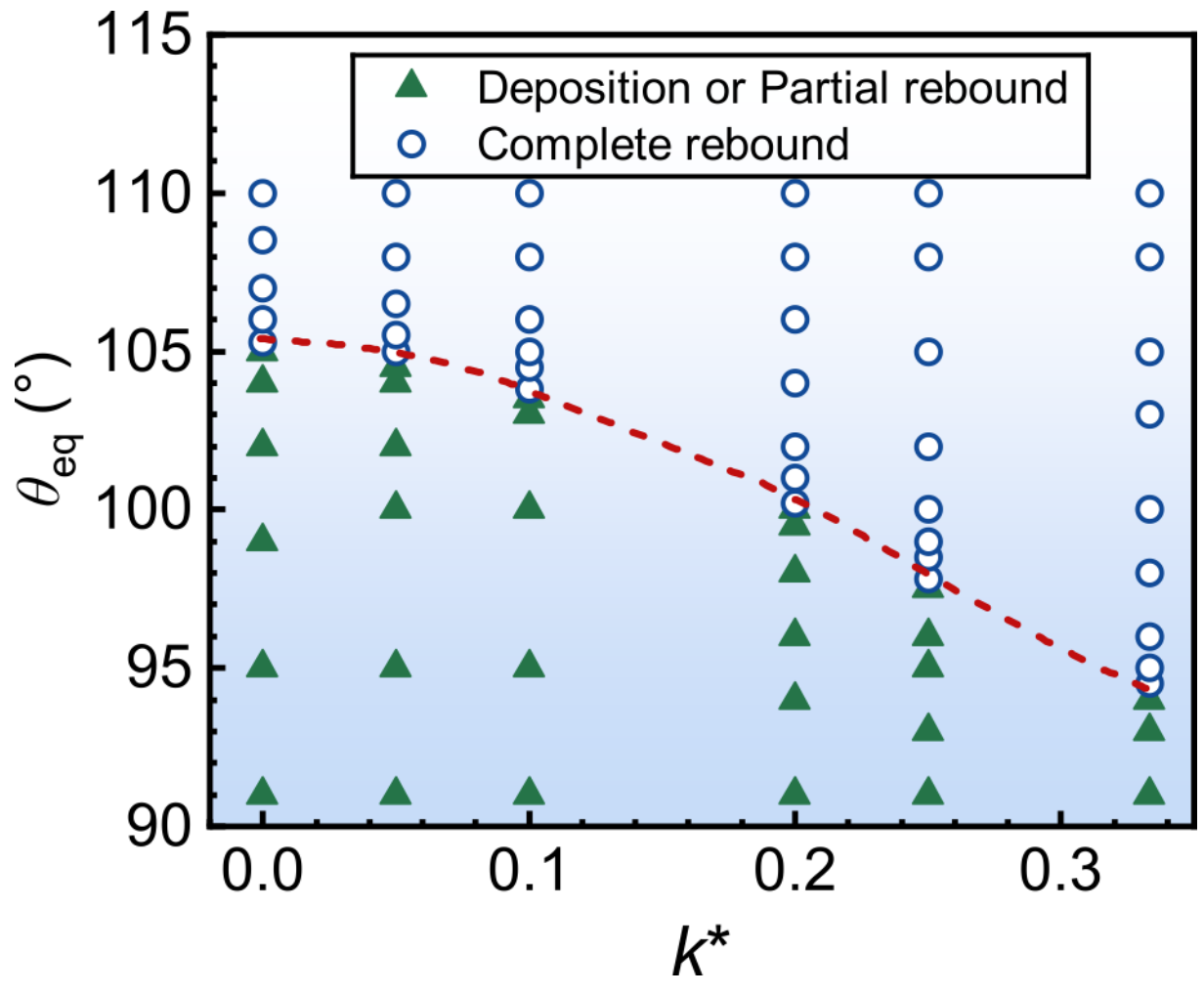

Figure 13. Variation of the critical contact angle for complete rebound mode with the dimensionless curvature. Here, $We$ = 27.7 and $Oh$ = 0.00237.

To further examine the influence of surface geometry on droplet impact dynamics, an additional comparison between cylindrical and spherical surfaces was performed under matched impact conditions, including the same Weber number, Ohnesorge number, characteristic curvature, and equilibrium contact angle. Figure 14 compares the temporal evolution of the total impact force for the two geometries. Both cases exhibit two distinct force peaks, corresponding to the initial inertial impact and the subsequent retraction process. However, the cylindrical surface exhibits a larger first impact-force peak and a shorter contact time than the spherical surface. Differences are also observed in the magnitude and occurrence time of the second force peak. These results indicate that the surface geometry affects the momentum transfer during initial impact and the liquid retraction dynamics. Unlike a cylindrical surface, which has only one nonzero principal curvature, a spherical surface has two nonzero principal curvatures. The observed differences suggest that the geometrical characteristics of the target surface play a role in the transient force response.

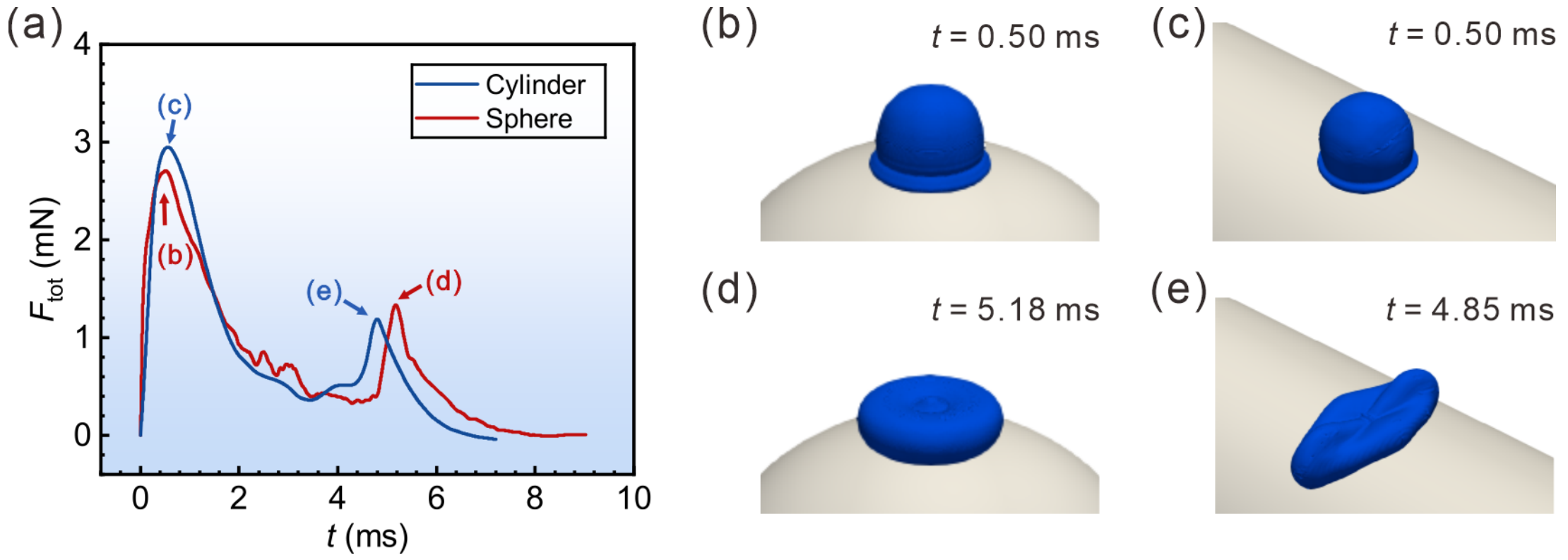

Figure. 14. Comparison of impact forces and droplet morphologies for cylindrical and spherical surfaces with equal curvature: (a) temporal evolution of the total impact force; (b) and (c) are droplet morphologies at the first force peak for the sphere and cylinder, respectively; (d) and (e) are droplet morphologies at the second force peak for the sphere and cylinder, respectively. Here, $\theta_{eq}$ = 165°, $We$ = 27.7, $Oh$ = 0.00237, $k^*$ = 0.2.

### 3.4 Effect of the Weber number

Throughout the impact process, the Weber number $We = \rho_l U_i^2 D_d / \sigma$ is a key dimensionless parameter that quantifies the competition between inertial and capillary effects. It directly affects droplet deformation and momentum transfer. In this study, different $We$ are obtained by adjusting the droplet impact velocity. As shown in Fig. 15(a), the dimensionless contact time $t_c$ / $t_\sigma$ remains close to 2.5 under different Weber numbers. The full rebound process is mainly controlled by the inertial-capillary time-scale and is almost independent of $We$. Since $We$ is changed mainly by changing the initial droplet velocity, the time required to reach the maximum spreading length $t_{max}$ is normalized using the inertial time-scale $t_\rho = D_d / U_i$. Dividing both sides of Eq. (31) by $t_\rho$ gives

$$t_{max} / t_\rho \sim We^{1/2} \tag{47}$$

Equation (47) is obtained by expressing the inertial-capillary timescale in terms of the inertial timescale. It therefore represents a change in time normalization rather than an independent scaling law. The numerical results are consistent with this expected timescale dependence. As shown in Fig. 15(b), an increase in *We* leads to a significant increase in $t_{\max} / t_{\rho}$. Figure 15(c) presents the variation in the maximum spreading length with the Weber number. It can be seen that the maximum spreading length on the spherical surface increases clearly with increasing *We*. This indicates that stronger impact inertia drives the liquid film to spread over a wider region along the spherical surface. There are two typical scaling laws governing the relationship between the maximum spreading length and *We*, corresponding to different regimes. If viscous dissipation and internal kinetic energy loss are neglected, and the initial kinetic energy of the droplet can be assumed to be fully converted into additional surface energy at the maximum spreading state, then the maximum spreading factor follows $D^*_{\text{s-max}} \sim We^{1/2}$ [71]. This relation corresponds to the purple dashed line in Fig. 15(c), and can be regarded as the upper limit under ideal energy conversion conditions. If the momentum-capillary balance and liquid volume conservation are considered, the maximum spreading length on a flat surface usually follows $D^*_{\text{s-max}} \sim We^{1/4}$ [71], which is represented by the red dashed line in Fig. 15(c). In this study, increasing spherical curvature promotes the surface-following spreading of the droplet, as evidenced by the increase in the maximum dimensionless spreading arc length $D^*_{\text{s-max}}$. The fitted result gives $D^*_{\text{s-max}} \sim We^{0.38}$, where the exponent lies between 1/4 and 1/2. This scaling indicates that the maximum spreading process of a droplet impacting a sphere does not fully follow the classical scaling for flat-wall impact. The spherical curvature enhances the overall spreading ability of the droplet. Therefore, the scaling exponent 0.38 obtained here is higher than the 1/4 value predicted by the inertial-capillary scaling, but is still lower than the ideal energy-conversion limit of 1/2. This suggests that viscous dissipation and internal flow losses still exist during spreading, and the initial kinetic energy cannot be completely converted into newly generated surface energy.

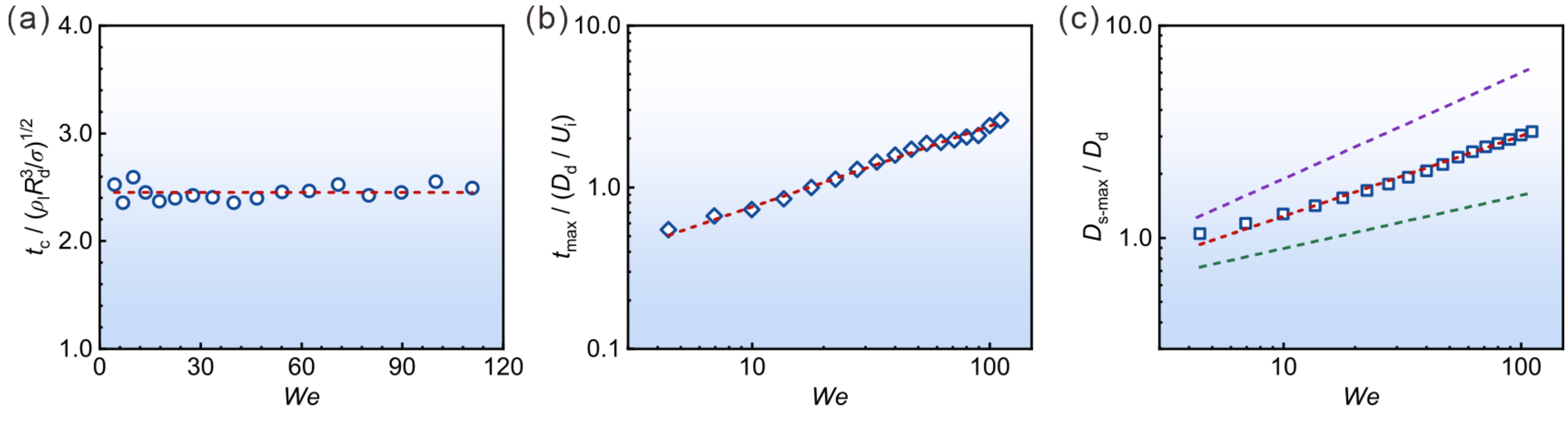

Figure 15. Effect of the Weber number on the kinematic parameters: (a) contact time $t_c$; (b) time to reach the maximum spreading length $t_{\max}$; (c) maximum spreading length $D_{\text{s-max}}$. Here, the contact time,

the time to maximum spreading, and the spreading length are normalized by the inertial-capillary time $t_\sigma = \left(\rho_l R_d^3/\sigma\right)^{1/2}$, the inertial time $t_\rho = D_d/U_i$ and the droplet diameter $D_d$, respectively. $\theta_{eq}$ = 165°, $Oh$ = 0.00237, $k^*$ = 0.2.

The effect of the Weber number on the impact force peaks and their occurrence times is shown in Fig. 16. As shown in Fig. 16(a), the occurrence time of the first impact force peak $t_{ft1}$ / $t_\sigma$ decreases significantly with increasing $We$. This is because the first peak forms during the initial impact stage of the droplet. Its occurrence time is mainly governed by the inertial impact time-scale. Therefore, the characteristic occurrence time of the first peak can be approximated as $t_{ft1} \sim D_d/U_i$. Thus, after normalizing $t_{ft1}$ by $t_\sigma$, the following scaling relation is obtained

$$t_{ft1}/t_\sigma \sim We^{-1/2} . \tag{48}$$

Similarly, Eq. (48) follows from the inertial timescale governing the initial impact stage. The relation reflects the expected dependence on the Weber number arising from the choice of normalization, rather than a newly established empirical correlation. This indicates that, as $We$ increases, the initial impact process becomes faster, and the first impact force peak appears earlier. In contrast, the occurrence time of the second peak $t_{ft2}$ / $t_\sigma$ increases slightly with increasing $We$, and then gradually stabilizes at approximately 1.4. This is because the second peak is mainly related to droplet retraction and central liquid convergence. Its occurrence time is not only controlled by the initial inertial time-scale, but is also affected by the preceding spreading range and the subsequent retraction dynamics. Therefore, as $We$ increases, the spreading range of the droplet becomes larger. The retraction and central convergence processes are then delayed, leading to a slight increase in the occurrence time of the second peak. The dimensionless first impact force peak $F_{t1}^*$ decreases with increasing $We$ and gradually approaches a stable value at high Weber numbers, as shown in Fig. 16(b). This is because the effect of surface tension cannot be neglected at low Weber numbers. At high Weber numbers, however, surface tension becomes negligible compared with inertia. The first impact force peak $F_{t1}$ can be expressed as the combined contribution of inertia and surface tension $F_{t1} = \beta_1\left(\rho_l U_i^2 D_d^2\right) + \beta_2\left(\sigma D_d\right)$, where $\beta_1$ and $\beta_2$ are constants [36]. The dimensionless first impact force peak $F_{t1}^*$ can then be written as

$$F_{t1}^* = \beta_1 + \beta_2 We^{-1} . \tag{49}$$

By fitting the data, we can obtain $F_{t1}^* = 0.60 + 2.11We^{-1}$. This indicates that $F_{t1}^*$ consists of an inertia-dominated term and a surface tension correction term. The constant term 0.60 corresponds to the inertial limit at high Weber numbers, while $2.11We^{-1}$ represents the contribution of surface tension at low Weber numbers. Similar functional relationships have been reported for droplet impacts on flat and cylindrical surfaces[36,45]. The common *We* dependence reflects the competition between inertia and surface tension, whereas the fitted coefficients differ among the three geometries. This difference arises because surface curvature modifies the redistribution of the incident droplet momentum, thereby affecting the inertia-dominated first impact force peak. Specifically, the inertia-dominated asymptotic coefficient follows the order $\beta_{1-\mathrm{flat}} > \beta_{1-\mathrm{cylinder}} > \beta_{1-\mathrm{sphere}}$. The variation of the dimensionless second impact force peak $F_{t2}^{*}$ with *We* is more complex, especially in the low Weber number range, as shown in Fig. 16(c). This behavior is related to the transition of retraction modes during the droplet retraction stage. At low Weber numbers, several retraction forms may occur during droplet impact, including Worthington jets, singular jets, and bubble entrapment [72]. These different retraction behaviors lead to a more complex variation in the second impact force peak.

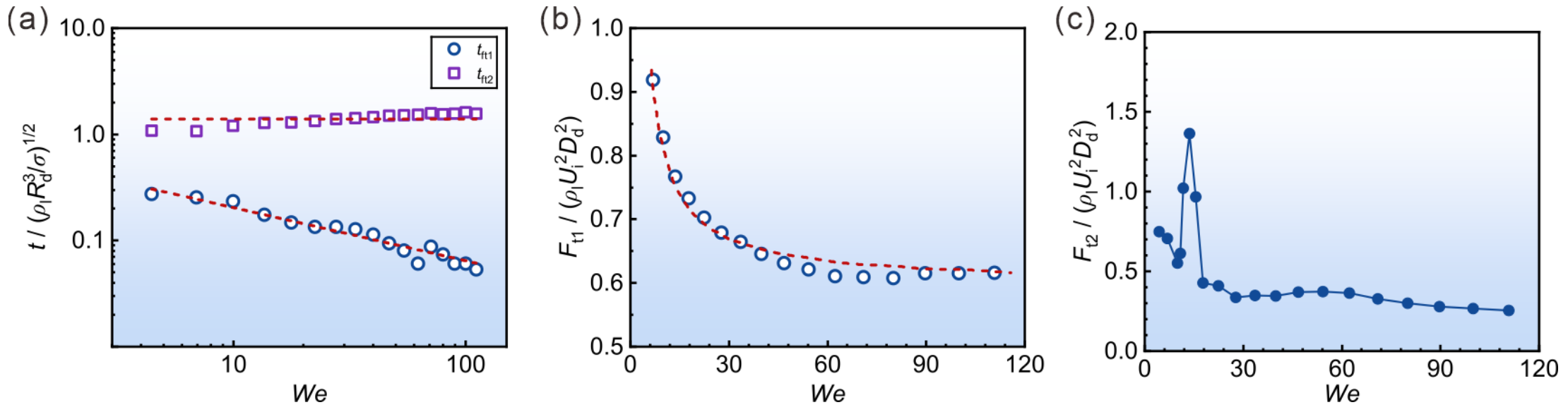

Figure 16. Effect of the Weber number on the dynamic parameters: (a) occurrence times of the first and second peaks, $t_{ft1}$ and $t_{ft2}$; (b) first impact force peak $F_{t1}$; (c) second impact force peak $F_{t2}$. Here, the time and force are normalized by the inertial-capillary time $t_{\sigma} = \left(\rho_l R_d^3/\sigma\right)^{1/2}$ and $\rho_l U_i^2 D_d^2$, respectively. $\theta_{eq} = 165°$, $Oh = 0.00237$, $k^{*} = 0.2$.

### 3.5 Effect of the Ohnesorge number

The Ohnesorge number $Oh = \mu_l / \sqrt{\rho_l D_d \sigma}$ is an important dimensionless parameter that characterizes the relative importance of viscous dissipation and capillarity during droplet impact. In this study, *Oh* is varied by changing the liquid viscosity. Since the liquid viscosity also enters the contact line capillary number in the modified Kistler model, varying Oh affects both viscous dissipation and the dynamic contact angle response. Yoon and Shin [24] similarly investigated droplet impact on spherical particles using a capillary number dependent dynamic contact angle model and identified viscous dissipation as a major factor controlling droplet spreading in the viscous regime. Therefore, although dynamic wetting also contributes to the observed *Oh*-dependent trends, viscous dissipation is expected to play an important role, particularly at higher *Oh*. At low *Oh*, viscous effects within the droplet are relatively weak. The retraction after spreading is mainly driven by surface tension, allowing the droplet to recoil and rebound rapidly. As *Oh* increases, viscous effects become increasingly important, and stronger viscous damping causes a larger fraction of the droplet's mechanical energy to be dissipated during retraction, even though the local velocity gradients may be attenuated. Consequently, less kinetic energy remains available to drive the recoil. This prolongs the time required for the droplet to retract following maximum spreading and subsequently detach from the spherical surface. Therefore, the dimensionless contact time $t_c$ / $t_\sigma$ gradually increases with increasing *Oh*, as shown in Fig. 17(a). The time to reach the maximum spreading length $t_{max}$ / $t_\sigma$ changes only slightly within the *Oh* range considered in this study, and increases slightly in the high *Oh* region, as shown in Fig. 17(b). This indicates that, when $Oh < 0.1$, the time to reach the maximum spreading state is still mainly determined by inertia. The effect of viscosity on this time $t_{max}$ / $t_\sigma$ is weaker than its effects on the spreading amplitude and the retraction process. It affects the maximum spreading size of the droplet, as shown in Fig. 17(c). The maximum spreading factor $D^*_{\text{s-max}}$ decreases monotonically with increasing *Oh*, following a power-law relation, $D^*_{\text{s-max}} \sim Oh^{-0.05}$. A larger *Oh* corresponds to stronger viscous dissipation. In this case, the initial kinetic energy of the droplet is more rapidly dissipated through internal viscous dissipation during spreading, instead of continuously driving the liquid film to extend along the spherical surface. Consequently, the maximum spreading extent of the droplet decreases. The fitted exponent of -0.05 indicates a relatively weak dependence of maximum spreading on viscosity within the investigated *Oh* range.

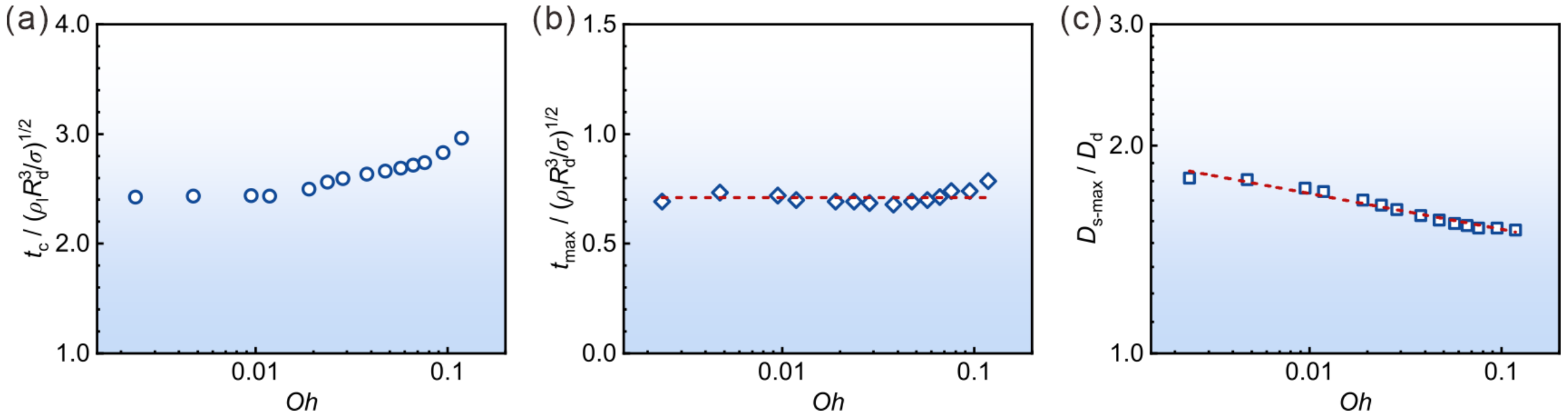

Figure 17. Effect of the Ohnesorge number on the kinematic parameters: (a) contact time $t_c$; (b) time to reach the maximum spreading length $t_{max}$; (c) maximum spreading length $D_{\text{s-max}}$. Here, the inertial-capillary time $t_\sigma = \left(\rho_l R_d^3/\sigma\right)^{1/2}$ and the droplet diameter $D_d$ are used to normalize time and length, respectively. $\theta_{eq} = 165°$, $We = 27.7$, $k^* = 0.2$.

The dimensionless first impact force peak $F^*_{t1}$ increases slightly with increasing *Oh*, and then gradually approaches a stable value at relatively high *Oh*, as shown in Fig. 18(a). As *Oh* increases, viscous effects within the droplet become more pronounced. The tangential spreading of the droplet during the initial impact stage is suppressed, and the momentum is more concentrated in the normal direction toward the wall. As a result, the dimensionless first peak increases slightly. However, the increase is limited, indicating that $Oh$ only has a weak effect on the initial impact force peak. Figure 18(b) shows that increasing $Oh$ has little influence on the occurrence time of the first peak. The value of $t_{\mathrm{ft1}} / t_\sigma$ remains almost unchanged. This indicates that the occurrence time of the first peak is mainly determined by the initial inertial impact process, rather than by viscous diffusion. Figure 18(c) shows that the dimensionless second impact force peak $F_{\mathrm{t2}}^*$ first increases and then decreases with increasing $Oh$, reaching a relatively large value near $Oh$ = 0.01. At low $Oh$, a moderate increase in droplet viscosity suppresses interfacial oscillations and irregular deformation, allowing the retracting liquid to converge more coherently toward the impact center. This more organized inward convergence enhances the upward momentum generated during retraction and consequently increases the second force peak. This leads to a larger second peak. However, when $Oh$ further increases, viscous dissipation becomes dominant. The retraction velocity of the droplet decreases significantly, and the rebound momentum is weakened. As a result, the second peak decreases rapidly.

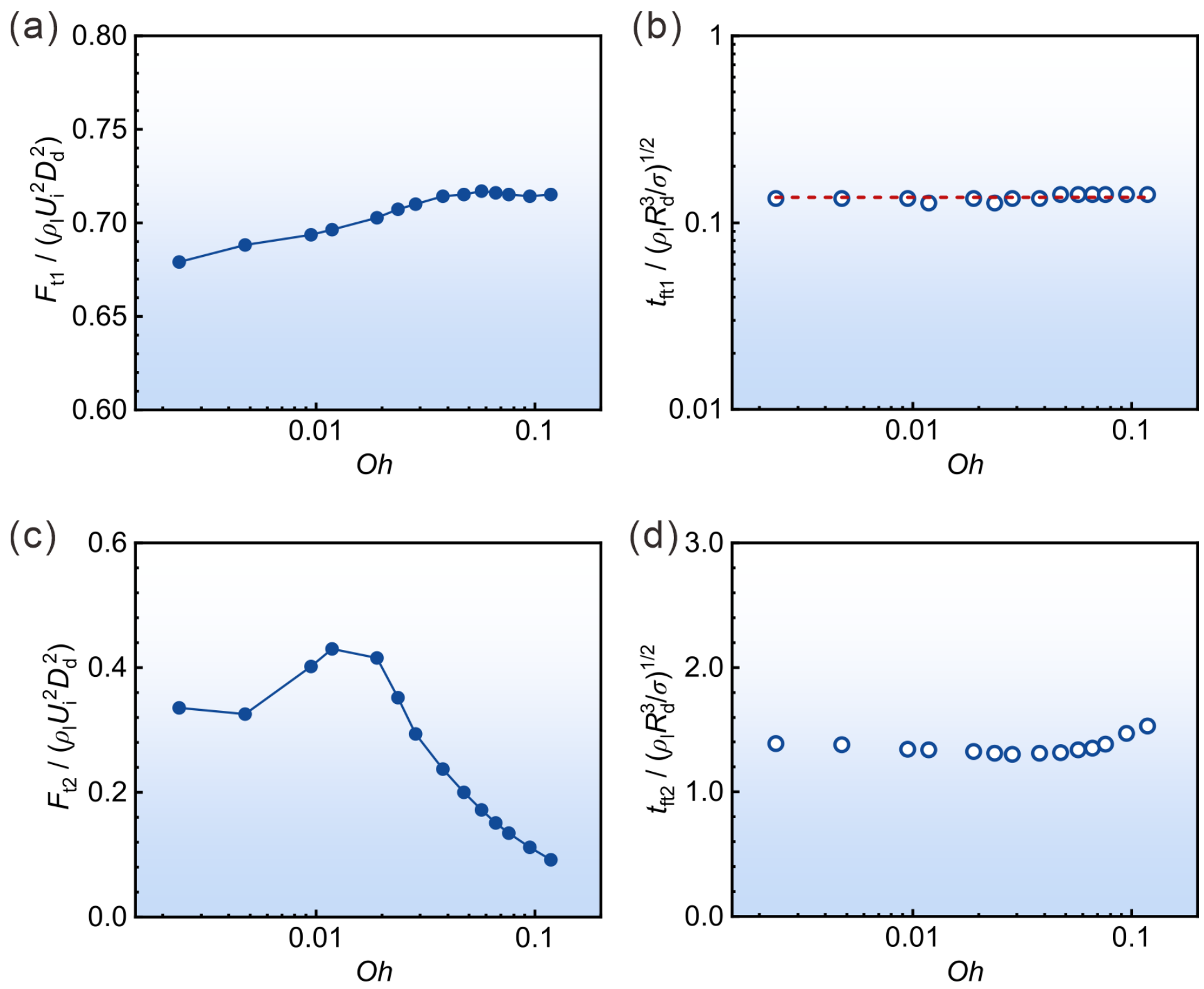

Figure 18. Effect of the Ohnesorge number on the dynamic parameters: (a) first impact force peak $F_{\mathrm{t1}}$; (b) occurrence time of the first peak $t_{\mathrm{ft1}}$; (c) second impact force peak $F_{\mathrm{t2}}$; (d) occurrence time of the second peak $t_{\mathrm{ft2}}$. Here, the time and force are normalized by the inertial-capillary time $t_\sigma = \left(\rho_\mathrm{l} R_\mathrm{d}^3 / \sigma\right)^{1/2}$ and $\rho_\mathrm{l} U_\mathrm{i}^2 D_\mathrm{d}^2$, respectively. $\theta_{\mathrm{eq}}$ = 165°, $We$ = 27.7, $k^*$ = 0.2.

It should be noted that the present parametric study investigates the effects of individual governing parameters while keeping the others fixed. Therefore, the empirical correlations obtained in Sections 3.2-3.5 are applicable to the corresponding investigated conditions and parameter ranges. Establishing a unified multi-parameter predictive model would require additional simulations considering the coupled effects of wettability, spherical curvature, Weber number, and Ohnesorge number, as well as the transitions among different impact regimes.

## 4. Conclusions

In this study, numerical simulations are performed to investigate the kinematic and transient force dynamics of droplets impacting spherical surfaces. The effects of wall wettability, sphere diameter, Weber number, and Ohnesorge number on droplet spreading, retraction, rebound, and wall impact force are analyzed. The results show that droplets impacting spherical surfaces mainly exhibit three typical modes: complete rebound, partial rebound, and complete deposition. In the complete rebound and partial rebound modes, the wall impact force shows two peaks. The first peak is mainly caused by the rapid reduction of the droplet momentum during the initial contact stage, while the second peak corresponds to the reaction force associated with the convergence and upward acceleration of the retracting liquid during rebound. Wall wettability significantly affects the maximum spreading length and the time required to reach maximum spreading. As the contact angle increases, both the maximum spreading length and the corresponding time decrease. In the complete rebound mode, the contact time decreases with increasing contact angle and can be expressed as $t_c/t_\sigma = C_1\left[\left(1-\cos\theta_{eq}\right)^{-1/2} - 1/\sqrt{2}\right] + C_2$, where $C_1$ and $C_2$ are constants. An increase in the dimensionless curvature increases both the maximum spreading length and the time required to reach maximum spreading. The dimensionless curvature changes the spreading morphology and the wall impact force response, but the contact time in the complete rebound mode remains nearly unchanged. Within the investigated range $0 \le k^* \le 0.33$, increasing the dimensionless curvature reduces the critical contact angle required for complete rebound. Increasing the Weber number significantly enhances droplet spreading, whereas the contact time is nearly independent of *We*. The occurrence time of the first impact force peak decreases with increasing *We*. The relationship between the dimensionless first impact force peak and *We* can be written as $F_{t1}^* = \beta_1 + \beta_2 We^{-1}$, where $\beta_1$ and $\beta_2$ are fitting constants. The variation of the dimensionless second impact force peak with *We* is more complex, and the fluctuations in the low *We* range are related to the transition of retraction modes. An increase in the Ohnesorge number causes a monotonic decrease in the maximum spreading length and a gradual increase in the contact time. The time required to reach maximum spreading and the occurrence time of the first impact force peak are not sensitive to *Oh*. The first impact force peak increases slightly with *Oh*, while the second peak is more sensitive to *Oh* and it first increases and then decreases. Although the present conclusions are restricted to axisymmetric impact conditions and the investigated parameter ranges, the present results contribute to a deeper understanding of the kinematic evolution and dynamic response mechanisms involved in droplet impact on spherical surfaces, and may provide useful guidance for spray cooling, anti-icing, droplet capture, and the design of curved functional surfaces.

## Supplementary Material

See the supplementary material for tests of computational-domain and mesh-size independence, validation of the transient impact force against experimental results, and verification of the impact force decomposition into pressure, viscous, and capillary contributions.

## Acknowledgements

This work was supported by the National Natural Science Foundation of China (Grant No. U25A20191).